\documentclass[aps,pra,reprint,amsmath,amssymb,,superscriptaddress]{revtex4-2}

\usepackage{graphicx}   
\usepackage{xcolor}     
\usepackage{soul}       
\usepackage{bm}         
\usepackage{url}        
\usepackage[
  colorlinks=true,
  linkcolor=blue,
  citecolor=blue,
  urlcolor=blue
]{hyperref}
\newif\ifcomments
\commentstrue            

\ifcomments
  \newcommand{\pcom}[1]{{\color{orange} [PG: #1]}}
  \newcommand{\ccom}[1]{{\color{red}    [CG: #1]}}
  \newcommand{\jcom}[1]{{\color{orange} [JL: #1]}}
  \newcommand{\acom}[1]{{\color{blue}   [AK: #1]}}
\else
  \newcommand{\pcom}[1]{}
  \newcommand{\ccom}[1]{}
  \newcommand{\jcom}[1]{}
  \newcommand{\acom}[1]{}
\fi

\begin{document}

\title{Long-range waveguide quantum electrodynamics with left-handed transmission lines}

\author{P. Goswami}
\affiliation{Department of Physics and Astronomy, Northwestern University, Evanston, IL 60208, USA}

\author{J. Liu}
\affiliation{Department of Physics and Astronomy, Northwestern University, Evanston, IL 60208, USA}

\author{C. A. Gonz\'{a}lez-Guti\'{e}rrez}
\affiliation{Department of Physics and Applied Physics, University of Massachusetts, Lowell, MA 01854, USA}
\affiliation{Instituto de Ciencias F\'{i}sicas, Universidad Nacional Aut\'{o}noma de M\'{e}xico, Cuernavaca 62210, Mexico}

\author{A. Kamal}
\email{archana.kamal@northwestern.edu}
\affiliation{Department of Physics and Astronomy, Northwestern University, Evanston, IL 60208, USA}
\affiliation{Department of Physics and Applied Physics, University of Massachusetts, Lowell, MA 01854, USA}

\date{\today}

\begin{abstract}
While engineering long-range light-matter interactions is the principal aim in waveguide-QED, ironically most of the building blocks rest on local short-range couplings, such as nearest-neighbor-coupled cavity arrays employed in canonical models. Here, we propose a waveguide-QED system with native long-range interactions, comprising a single emitter coupled to a left-handed transmission line (LHTL). Interestingly, the LHTL emulates a synthetic photonic lattice with a slow logarithmic decay of hopping amplitudes over a distance set entirely by the ratio of UV and IR cutoffs of line dispersion. Its intrinsic long-range nature manifests both in the properties of atom-photon bound and scattering states, which exhibit algebraic localization and accelerated photon propagation respectively. Using a method of `running exponents', we develop a unified picture connecting waveguide dispersion to bound state and light front profiles obtained in the strong long-range hopping regime. These results motivate how transmission lines can enable multi-qubit information processing with tunable-range interactions.
\end{abstract}
\maketitle
\section{Introduction}
%

Waveguide quantum electrodynamics has emerged as a new paradigm to explore cooperative effects in light-matter interactions and implement new quantum information functionalities based on itinerant photons, such as remote entanglement generation and quantum communication. Low dimensional waveguide-QED platforms, such as those based on superconducting qubits coupled to microwave waveguides (a.k.a transmission lines), are particularly appealing given their ease of control, high coupling efficiencies \cite{wQEDReview2023}, as well as the significant design freedom afforded by such systems -- all the way from tunable atoms with user-defined coupling modalities, to exotic waveguide geometries. A recent example is `giant' atom waveguide-QED where a single qubit can be coupled at multiple points to a transmission line in order to engineer effects such as nonexponential atomic decay \cite{Andersson2019} and directional photon transmission \cite{Kannan2023}. It is worthwhile to note that almost all such variations and demonstrations till date are based on right-handed transmission lines (RHTLs), with the design focus being on the qubit or its coupling to the transmission line. Inspired by ideas from optical metamaterials, there has been increasing interest in exploring opportunities offered  by engineering the transmission line or waveguide instead \cite{Indrajeet2020,Zhang_science2023} that can enable almost a `brass tacks' approach to modifying the spectral properties of the photonic continuum. In this vein, one-dimensional left-handed transmission lines (LHTLs) present an intriguing alternative and a perfect dual to their conventional right-handed counterparts, with the roles of inductors and capacitors along the line interchanged. While such a change may seem superficially minimal at first glance, it constitutes a highly non-trivial modification of the line dispersion, as evidenced by the new physical effects already reported in LHTL systems, namely particle production out of the quantum vacuum \cite{FerreriPRR2024}, generation of highly entangled states \cite{EggerWilhelm2013}, and simulation of interesting quantum optical and thermodynamic effects in composite LH/RHTL systems \cite{FerreriPRApp2025}. 
\par
Of particular importance in waveguide-QED are the entangled states known as atom-photon bound states, which are non-radiative states arising due to photonic band gaps. These states lead to long-lived, spatially-localized excitations due to vanishing group velocities near the band edge and can be leveraged to implement strong photonic or multi-emitter interactions. Recently there has been renewed interest in waveguide engineering, via periodic loading \cite{Mirhosseini2018}, staggered hoppings \cite{Kim_TopologicalMetamaterialWaveguide2021} or impedance engineering \cite{Scigliuzzo2022}, to extend the range of localized bound-state wavefunction. Nonetheless, all these approaches focus almost exclusively on tweaking standard waveguide models based on coupled-cavity arrays with quadratic or cosine-like dispersion relations, and report an increase in the localization length while maintaining exponential bound-state profiles. In contrast, here we report how TL-based waveguide-QED systems exhibit functionally distinct algebraic localization, with a polynomial fall off with distance from the emitter. This is one of the central contributions of this paper and indicates how TL-based waveguide-QED provides a qualitatively new means to design long-range interactions in spin-boson and quantum impurity models. 
\par
Specifically, LHTL waveguide with its inverted dispersion maps to an unusual lattice hosting power-law hoppings with a distance-dependent exponent. This makes it a unique platform for emulating photonic lattices with strong long-range, weak long-range and short-range hoppings in the same system sampled at different length scales. Using this mapping as a backdrop, we establish direct connections between bound-state and scattering properties across these regimes, including appearance of algebraic localization and accelerated light cones due to strong long-range hoppings -- a regime hitherto inaccessible to waveguide QED. Crucially, though perhaps unsurprisingly, this complexity also reflects in enhanced non-Markovianity in dynamics of the emitter coupled to an LHTL, with relaxation times that can even decrease with increase in emitter frequency in contrast to the behavior observed in the presence of usual Ohmic baths. 
%
\section*{Results}
%
\subsection*{Left-handed transmission-line waveguide-QED platform}
%
\begin{figure}[t!]
    \centering
    \includegraphics[width=1\linewidth]{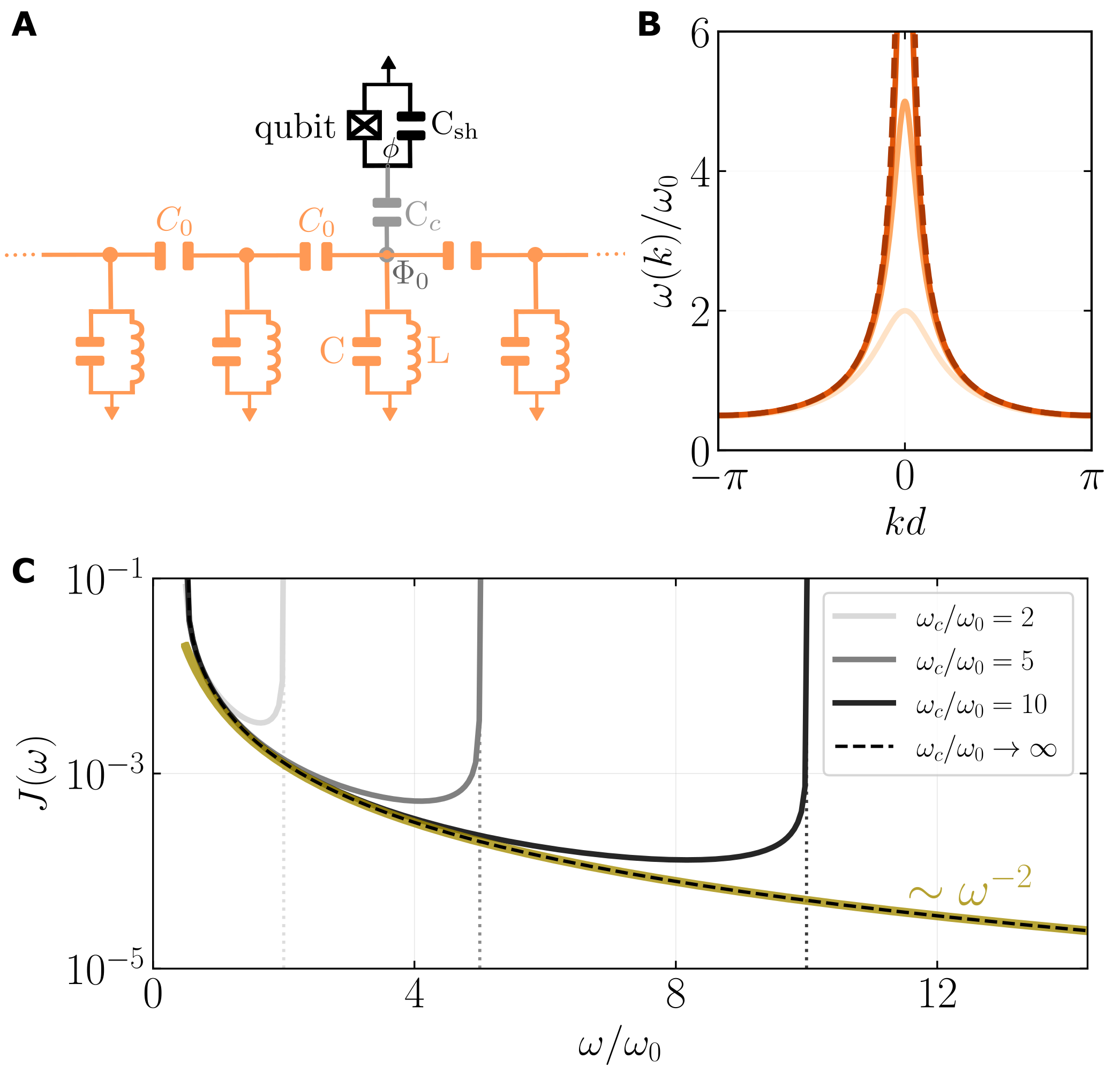}
    \caption{\textbf{Left-handed waveguide-QED.} \textbf{A} Circuit representation for a transmon qubit capacitively coupled to a LHTL. \textbf{B} LHTL dispersion for different values of UV cutoff $\omega_c$. \textbf{C} Spectral density seen by the qubit for the same set of cutoff frequencies.}
    \label{fig: Left-handed waveguide-QED architecture}
\end{figure}
The left-handed waveguide-QED system we consider here consists of a two-level emitter coupled locally to a left-handed transmission line (LHTL); Fig.\ref{fig: Left-handed waveguide-QED architecture}A shows a superconducting circuit-based realization of such a system with the two-level emitter shown as a superconducting transmon qubit, coupled capacitively to an LHTL depicted using the equivalent telegrapher model for microwave transmissions lines \cite{FerreriPRR2024}. The LHTL dispersion relation is given by
\begin{equation}
    \omega_k = \frac{\omega_0 \omega_c}{\sqrt{\omega_0^2 + 2\omega_c^2(1-\cos(kd))}},
\end{equation}
where $d$ denotes the inter unit-cell separation (set to unity in dimensionless units), $\omega_{0} = (LC_{0})^{-1/2}$. The left-handedness of the transmission line arises from the phase and group velocities being opposite in sign ($v_p v_g < 0$) as evident from the dispersion plotted in Fig.\ref{fig: Left-handed waveguide-QED architecture}B. It is worth noting that we have included a capacitance to ground in the LHTL model presented here; besides being relevant to realistic experimental realizations, this also imposes a finite UV cutoff $\omega_{c} = (LC)^{-1/2}$ on the LHTL passband. For infinite UV-cutoff ($\omega_c \to \infty$), the dispersion reduces to the form $\omega_k = \omega_0/2|\sin(k/2)|$ with has a pole at $k = 0$. Introducing a finite capacitance to ground regularizes this divergence without affecting the handedness of the line.
\par
Starting from the full circuit Lagrangian, we obtain the standard waveguide-QED Hamiltonian in the single-excitation regime under rotating-wave approximation (RWA),
\begin{equation}
    H=\Delta\,\sigma^+\sigma^- + \sum_{k = 1}^N \omega_k a_k^\dagger a_k+\sum_{k = 1}^N\,(g_k\; \sigma^+ a_k+ \text{h.c.}), 
\end{equation} 
where $\Delta$ is the emitter frequency, $\{g_{k}\}$ denote the momentum-dependent emitter-waveguide coupling constants, and $N$ is the number of unit cells in the transmission line. Here $k$ indexes the wave-vector number and associated operators defined in the momentum basis (see Supplementary Materials). The inverted dispersion of LHTL presents an unconventional spectral profile to the emitter, $J(\omega) = 2\pi \sum_{k} |g_k|^2\,\delta(\omega - \omega_k)$,
\begin{subequations}\label{eq: Spectral function LHTL flat gk}
\begin{align}
    J(\omega) &= \frac{2g^2\omega_0^2}{\omega^3 \sqrt{1 - \left[ 1 - \frac{\omega_0^2}{2}\left( \frac{1}{\omega^2} - \frac{1}{\omega_c^2}\right) \right]^2}}, \\
    &{\approx}\frac{2g^2 \omega_0}{\omega^2}, \quad {\rm for } \; \omega_0 \ll \omega \ll \omega_c. \label{subeq: Deep in band spectrum}
    \end{align}
\end{subequations}
\begin{figure*}[t!]
    \centering
    \includegraphics[width=1\linewidth]{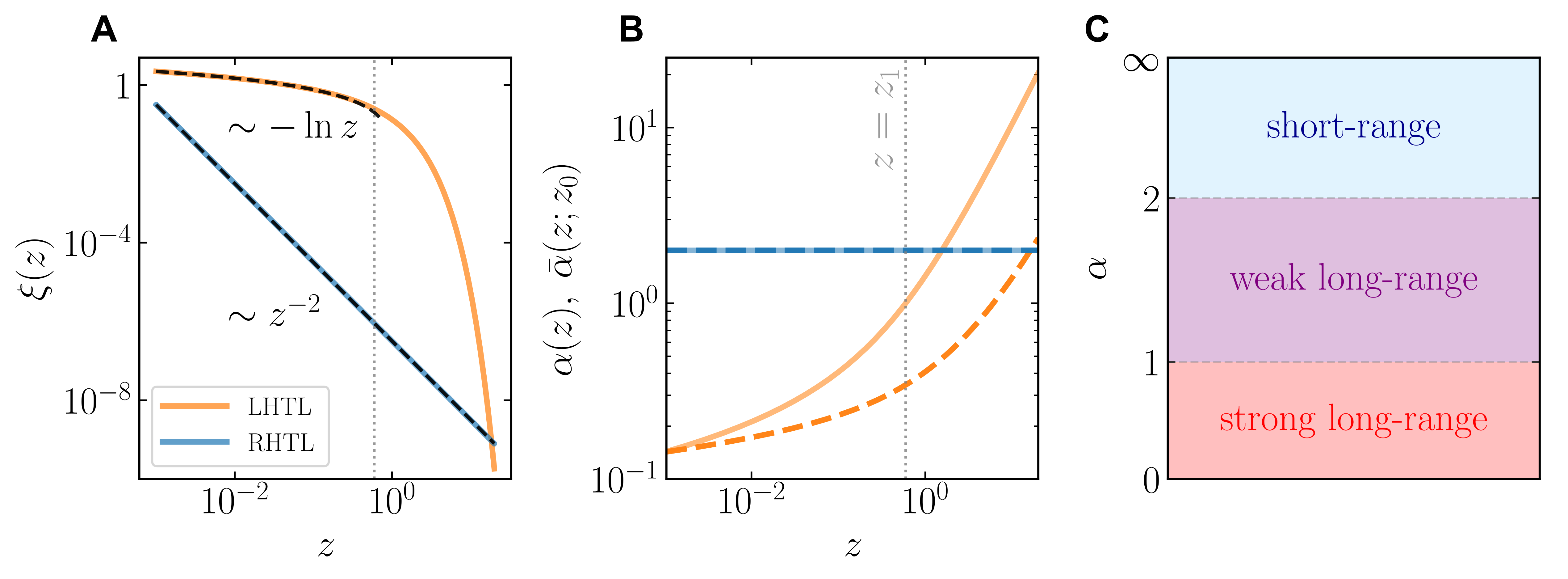}
    \caption{\textbf{Hopping amplitudes for transmission lines.} \textbf{A} LHTL maps to a tight-binding photonic lattice with logarithmic fall off of hopping amplitudes $\xi(z)$ for $z<1$, and exponential fall off for $z>1$ ($z \equiv n/n_\star$), while RHTL maps to a power-law hopping network with a constant exponent $\alpha = 2$. \textbf{B} Position-dependent local (solid) and global (dashed) hopping exponents for LHTL and RHTL obtained using the running exponent method. \textbf{C} Long-range regimes for 1D power-law hopping network, with $\alpha \to \infty$ corresponding to nearest-neighbor models.}
    \label{fig: LH and RH hopping network}
\end{figure*}%
The behavior of the spectral function $J(\omega)$ with increasing UV cutoff ($\omega_{c}$) is shown in Fig.~\ref{fig: Left-handed waveguide-QED architecture}C. It exhibits square-root van Hove singularities at the two band edges $\omega_{\rm IR} = \omega_0\omega_c/\sqrt{\omega_0^2 + 4\omega_c^2}$ and $\omega_{\rm UV} = \omega_c$, though the UV and IR frequency cutoffs now appear at long ($k = 0$) and short ($k = \pi$) wavelengths, respectively, due to inverted LHTL dispersion. As shown in Eq.~\ref{subeq: Deep in band spectrum}, away from the band-edges, LHTL presents a Brownian spectrum $J(\omega) \sim 1/\omega^2$, making it natively non-Markovian. This is in sharp contrast to conventional right-handed systems and coupled cavity-arrays that almost always behave as Markovian environments due to their spectral function being flat away from the edges, and which typically rely on non-local emitter-waveguide couplings (giant-atom) or time-delay geometries to engineer non-Markovian effects. Further, even for non-Markovian noise, $J(\omega)\propto\omega^{s}$ with $s=-2$ is unusual from the perspective of spin-boson physics where studies on non-Ohmic spectra have been restricted to sub-Ohmic baths with $0<s<1$. While spin-boson models with inverted spectral densities ($s<0$) have recently been explored in the context of spin dephasing \cite{Otterpohl2025}, similar studies for finite-frequency processes such as spin relaxation have not been pursued to the best of our knowledge. Thus, transmission line-based waveguide-QED systems provide an interesting and experimentally-viable alternative to explore exotic spin-boson models with tailored spectral densities.
%
\par
Note that in Eq.~\ref{eq: Spectral function LHTL flat gk}, we have assumed a `flat' $g_{k} \equiv g/\sqrt{N}\;\forall\; k$ for analytical tractability. In the following sections, we will continue to use this simple coupling profile to study both bound- and scattering-state properties, since that allows highlighting the salient features due to inverted LHTL dispersion alone. Later, we will include the full momentum dependence of emitter-waveguide coupling, though this would not significantly alter the qualitative features of the results obtained with $g$; for instance, the emitter-photon bound state exhibits algebraic localization in both cases though the exponent is modified by the coupling profile $g_{k}$. 

\subsection*{Beyond power-law hoppings}
%
Starting with seminal work of Bak, Tang and Wiesenfeld \cite{BakTangWiesenfeld}, there have been several works which have provided tantalizing pointers on how spatially extended modes can lead to $1/\omega^{s}$ spectrum. Given the Brownian spectrum derived in the previous section, it is natural to expect that photonic correlations in LHTL-type systems must be long-range. To concretize this notion, we develop a photonic tight-binding lattice model for LHTL in real space, $H_{{\rm LH}} = \sum_{m\in\mathbb{Z}}\sum_{n\in\mathbb{Z}} \xi_n\, a_{m+n}^\dagger a_m$
where the hopping amplitudes $\xi_n$ are described by a modified Bessel function of second kind,
\begin{equation}\label{eq: LHTL hopping}
\xi_n \simeq \frac{\omega_0}{\pi}\, K_0\!\left(\frac{n\,\omega_0}{\omega_c}\right) \simeq \frac{\omega_0}{\pi}\, K_0\!\left(\frac{n}{n_\star}\right),\;\;
n_\star \equiv \frac{\omega_\text{UV}}{2\omega_\text{IR}}.
\end{equation}
The standard asymptotes of $K_0(z)$ give
\begin{equation}\label{eq: Asymptotes of K0}
    \xi(z) \simeq \frac{\omega_0}{\pi}
\begin{cases}
-\ln\!\big(\tfrac{z}{2}\big)-\gamma_E, & z\ll 1 \quad (1\ll n\ll n_\star),\\[6pt]
\displaystyle \sqrt{\frac{\pi}{2}}\,\frac{e^{-z}}{\sqrt{z}}, & z\gg 1 \quad (n\gg n_\star).
\end{cases}
\end{equation}
where, $z\equiv n/n_\star$,
and $\gamma_E \approx 0.577$ is the Euler's constant. The hopping amplitude profile of the LHTL exhibits a crossover from logarithmic to exponential falloff around $z = 1$ ($n = n_\star$), at a length scale entirely determined by the ratio of UV and IR cutoff frequencies (Fig.~\ref{fig: LH and RH hopping network}A). This ``ultra" long-range nature of hoppings in LHTL is further validated by the fact that a nearest-neighbor, or even the next-nearest-neighbor, real-space truncation of hoppings fails to reproduce the correct dispersion relation for LHTL (see Supplementary Materials). In contrast, the hopping amplitude profile for RHTL follows a power law, $\xi^\text{RH}(z) = (\omega_0^3/\pi \omega_c^2)\,z^{-2}$, with no crossover length scale. This is consistent with RHTL dispersion being gapless ($\omega_\text{IR}^\text{RH} = 0$), leading to $n_\star \rightarrow \infty$.
%
\par
In order to enable a direct comparison of hopping amplitudes in Eq.~(\ref{eq: Asymptotes of K0}) with standard long-range models with power-law interactions, we parameterize the spatial profile of hoppings with a \emph{scale-dependent} or ``running'' power-law exponent, $\xi(z) \sim z^{-f(z)}$. Applying this construction to left- and right-handed transmission lines, with the hopping profiles $\xi^\text{LH}(z) = (\omega_0/\pi)\, K_0(z)$, and $\xi^\text{RH}(z) = (\omega_0^3/\pi \omega_c^2)\,z^{-2}$, we obtain,
\begin{subequations}
    \begin{align}
    & \alpha^{\rm LH}(z)=z\,\frac{K_1(z)}{K_0(z)},\;\;
\bar{\alpha}^{\rm LH}(z;z_0)=\frac{\ln\!\big[K_0(z_0)/K_0(z)\big]}{\ln(z/z_0)},
\label{eq:alpha-LH}\\
& \alpha^\text{RH}(z) = \bar{\alpha}^\text{RH}(z,z_0) = 2. \label{eq:alpha-RH}
    \end{align}
\end{subequations}
 In the running-exponent language, the RHTL realizes a trivial, scale-invariant $1/z^2$ hopping network, whereas the LHTL exhibits genuine scale-dependent hopping amplitudes, see Fig.~\ref{fig: LH and RH hopping network}B. From Eq.~\ref{eq:alpha-LH}, we find a well-defined boundary at $z_1 \simeq 0.60$ where $\alpha(z_1) = 1$, which separates strong ($\alpha < 1$) and weak ($\alpha>1$) long-range hopping. In 1D, $\alpha = 1$ coincides with the integrability threshold of power-law tails \cite{Defenu_LR_Interacting_Systems_2023}. From Eq.~\ref{eq:alpha-LH}, we find well-defined boundaries at $z_1 \simeq 0.60$ and $z_2 \simeq 1.55$ where $\alpha(z_{1,2}) = 1,2$, which separates strong ($\alpha < 1$) to weak ($1<\alpha<2$) long-range hopping regime, and weak long-range to short-range hopping regimes ($\alpha>2$), respectively.
\par
The appearance of $n_\star$ as the relevant crossover scale underscores how LHTL waveguide-QED provides a crucial insight into relevant interaction scale for multimode systems. Past works have explored hybrid left/right-handed transmission lines in circuit-QED waveguide setups, and identified $\omega_\text{IR}^{-1}$ as the relevant interaction length scale by mapping the spin-boson model to an effective 1D Ising system \cite{EggerWilhelm2013}. However, as highlighted by LHTL hopping profiles, Eq.~\ref{eq: LHTL hopping}, the relevant length scale is $n_\star$ which is a ratio of UV and IR frequencies. Interestingly, this implies that interaction scale in LHTL-based waveguide networks can be pushed to infinity by pushing the waveguide UV cutoff $\omega_{c}$ higher even in the presence of a finite IR cutoff.
\begin{figure*}[ht!]
    \centering
    \includegraphics[width=1.0\linewidth]{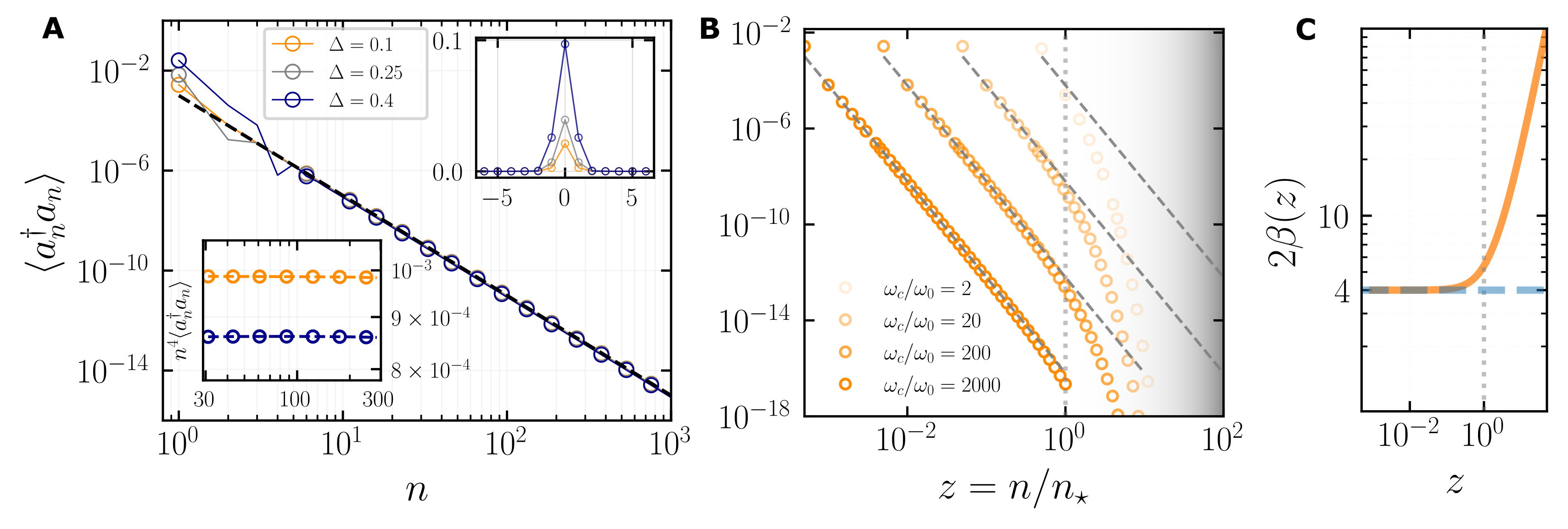}
    \caption{\textbf{Algebraic bound-states in LH waveguide-QED.} \textbf{A} Spatial profile of bound state for $n_\star \to \infty$, for different emitter frequencies ($g/\omega_0 = 0.1$) shown in different color markers. The dashed black line shows the fit to $1/n^4$ power-law. The top-right inset shows the photon intensity profiles near the qubit location for the three cases. The lower-left inset shows a slightly different asymptote values for two different emitter frequencies (detunings). \textbf{B} Spatial profiles of bound state, for different values of $\omega_{\rm UV} = \omega_{c}$, each showing the exponential decay in the region $z > 1 (n> n_{\star})$ (indicated with the gray gradient on the right). \textbf{C} Position-dependent local bound state exponent, $2\beta(z)$, for left-handed (solid-orange) and right-handed (dashed-blue) waveguide-QED systems, calculated using the running exponent method.}
    \label{fig: Algebraic bound-state crossover}
\end{figure*}
%
%
\subsection*{Algebraic localization and crossover}
\label{sec: Qubit-Photon Bound state}
%
Band edges in the spectral function $J(\omega)$ are essential for realizing photonic bound states when the emitter is tuned \emph{outside} the band. In the absence of any waveguide modes resonant with the emitter, spontaneous emission is suppressed and any initial excitation of the emitter instead hybridizes into a dressed eigenstate (a.k.a. bound-state) of a photon shared between emitter and waveguide, with the photonic part of the bound state wavefunction residing in the waveguide due to coupling between the emitter and the band-edge(s) (essentially the modes with $v_g$ near zero). In all waveguide-QED platforms studied to date, this bound-state is exponentially localized around the emitter\cite{wQEDReview2023}. In contrast, the long range nature of left-handed waveguide leads to a qualitatively different nature of localization leading to \emph{extended} bound-states as shown in Figure~\ref{fig: Algebraic bound-state crossover}B.
Notably, the LHTL bound state exhibits an \emph{algebraic} (power-law) tail with quartic fall-off with distance from the emitter (Fig.~\ref{fig: Algebraic bound-state crossover}A), which persists up to $n \ll n_\star$, and modifies to an Ornstein-Zernike form with an \emph{exponential} tail for
$n\!\gg n_\star$,
\begin{equation}\label{eq: Algebraic Bound-state LHTL}
    \langle a_n^\dagger a_n \rangle \simeq \frac{g^2|C_e|^2}{\pi^2 \omega_0^2}\begin{cases}
    \dfrac{1}{n^4}, &\quad (1\ll n\ll n_\star),\vspace{3mm}\\ 
    \left(\dfrac{\pi}{2 n_\star} \right)\,\dfrac{e^{-2n/n_\star}}{n^3}, &\quad (n\gg n_\star),
    \end{cases}
\end{equation}
Here we have used $K_1(z) \approx 1/z$ for $z \ll1$. The algebraic localization is a direct consequence of the intrinsic long-range couplings in the LHTL; this is clear from the qualitative change in bound state profile across $n_\star=\omega_\text{UV}/2\omega_\text{IR}$ when couplings crossover from strong long-range ($n \ll n_\star$) to short-range ($n \gg n_\star$). Recently, similar power-law localization has been reported for a qubit edge-coupled to 2D honeycomb lattice with nearest neighbor interactions \cite{EmergentCavityQEDARXIV2025} where range of interaction is determined by lattice anisotropy. 
\par
We establish a direct connection between bound-state profile and the underlying hopping network by deriving an analytic relation between the the local hopping exponent, $\alpha (z)$ (see Eq.~\ref{eq:alpha-LH}) and the local bound-state exponent, \(\beta(z)\equiv -\,d\ln|\phi_z|/d\ln z\) with $\phi(z)$ being the un-normalized bound-state amplitude, as
\begin{equation}\label{eq: BS envelope local exponent}
    \beta^\text{LH}(z)=2 + \frac{z\,K_0(z)}{K_1(z)} = 2+\frac{z^{2}}{\alpha^\text{LH}(z)},
\end{equation}
Here we have introduced a position-dependent exponent \(2\beta(z)\) that allows to describe a generic bound-state profile; as shown in Fig.~\ref{fig: Algebraic bound-state crossover}C, it equals \(4\) for \(z<1\) and scales approximately linearly with distance beyond $z = 1$. A constant local exponent corresponds to a pure power-law behavior while a linear increase with distance implies an exponential fall off. 

Since the term \(z^{2}/\alpha(z)\) is strictly positive for any finite \(n_\star\) (i.e., since \(\alpha(z)>0\)), the spatial decay of bound-state amplitude is never slower than \(n^{-4}\). Notably, the crossover point \(z=1\) $(n = n_{\star})$ lies between \(z_1\approx 0.60\) and \(z_2\approx 1.55\), the distances governing the crossover from strong- to weak long-range hopping and from weak long-range to short-range hopping in 1D systems respectively, i.e. \(\alpha(z_{1,2})=1,2\), \cite{Defenu_LR_Interacting_Systems_2023}. 
\par 
For comparison, we also show the bound state exponent obtained for the conventional right-handed waveguide-QED system in Fig.~\ref{fig: Algebraic bound-state crossover}C (see Supplementary Materials). In this case, the dispersion is gapless and linear for $k \to 0$, i.e., $\omega_k \simeq \omega_0 |k|$. For $\omega_c > \omega_0$, we find, $\xi^\text{RH}(n) \simeq \omega_0/\pi n^2$ and $|\phi_n|^2 = (\omega_0^2 g^2/\pi^2 E_\text{bound}^4) \;n^{-4}$ in the limit $n \gg 1$, showing that the resulting hopping network and the qubit-photon bound-state profile are essentially insensitive to $\omega_{\text UV}$. This is not entirely surprising since the system always operates in the $n_{\star} \to \infty$ limit since $\omega^\text{RH}_{\rm IR} =0$.  Thus, while the RHTL also exhibits algebraic bound-state profiles, it does so with constant exponents, $\alpha^\text{RH}(z) = 2$ and $\beta^\text{RH}(z) = 2$ with no crossover (see Fig.~\ref{fig: Algebraic bound-state crossover}C). 
\par
It is instructive to note that the standard coupled cavity-array picture of waveguide-QED with short-range or nearest-neighbor hoppings, corresponds to the  ``bad" waveguide limit of LHTL, see table.~\ref{tab:exactg_k}. Specifically, this is realized when $\omega_c \ll \omega_0$ or equivalently $C \gg C_0$, leading to an almost flat dispersion and suppressed hopping between neighboring sites (cavities). In this regime, we reproduce the familiar \emph{exponentially localized} bound-state profiles obtained for nearest-neighbor couplings.
\par
In addition to the $n_{\star}$-controlled crossover, the bound state has a short-range exponential core around the qubit, with the characteristic length scale $\lambda_\text{core} \propto \delta^{-1/2}$, where $\delta = \omega_\text{IR} - E_\text{bound}$ denotes the detuning of the (dressed) emitter frequency from the band edge \cite{ferreira2020,Bello2019} (see Supplementary materials). Such frequency-tunable bound state cores, with emitter parked very close to the band-edge, have often been proposed as a mechanism for implementing ``long-range" qubit-qubit interactions \cite{Zhang_science2023, Douglas2015-wc,Kim_TopologicalMetamaterialWaveguide2021, Scigliuzzo2022}. The extended bound states reported in our work present a qualitatively distinct paradigm for realizing long-range interactions both in terms of the functional form of the bound state (algebraic vs exponential), as well as the interaction length scale being independent of the emitter frequency and solely determined by the waveguide properties (specifically, $n_{\star} = \omega_{\rm UV}/2\omega_{\rm IR})$. From a practical design standpoint, extending the range of bound state via tuning of emitter frequency leads to a fine-tuning problem since a small detuning $\delta$ also makes the emitter susceptible to be in accidental resonance with modes near the band-edge especially due to coupling-induced Lamb shifts \cite{Zhang_science2023}.
%
\subsection*{Non-Markovianity and Accelerated Light Cones}
\label{sec: Exact Dynamics}
%
In this section, we investigate the effect of native long-range hoppings in the LHTL on the dynamics of scattered photons and qubit population. In the context of former, we will see how LHTL waveguide-QED system leads to unconventional light cones with position-dependent exponents, while for the latter it leads to non-Markovian decay for an emitter parked inside the continuum ($\omega_{\rm IR} < \Delta < \omega_{\rm UV}$). 
%
\subsubsection*{Non-Markovian Qubit Dynamics}
%
%
\begin{figure*}[!t] 
    \centering 
    \includegraphics[width=1.0\linewidth]{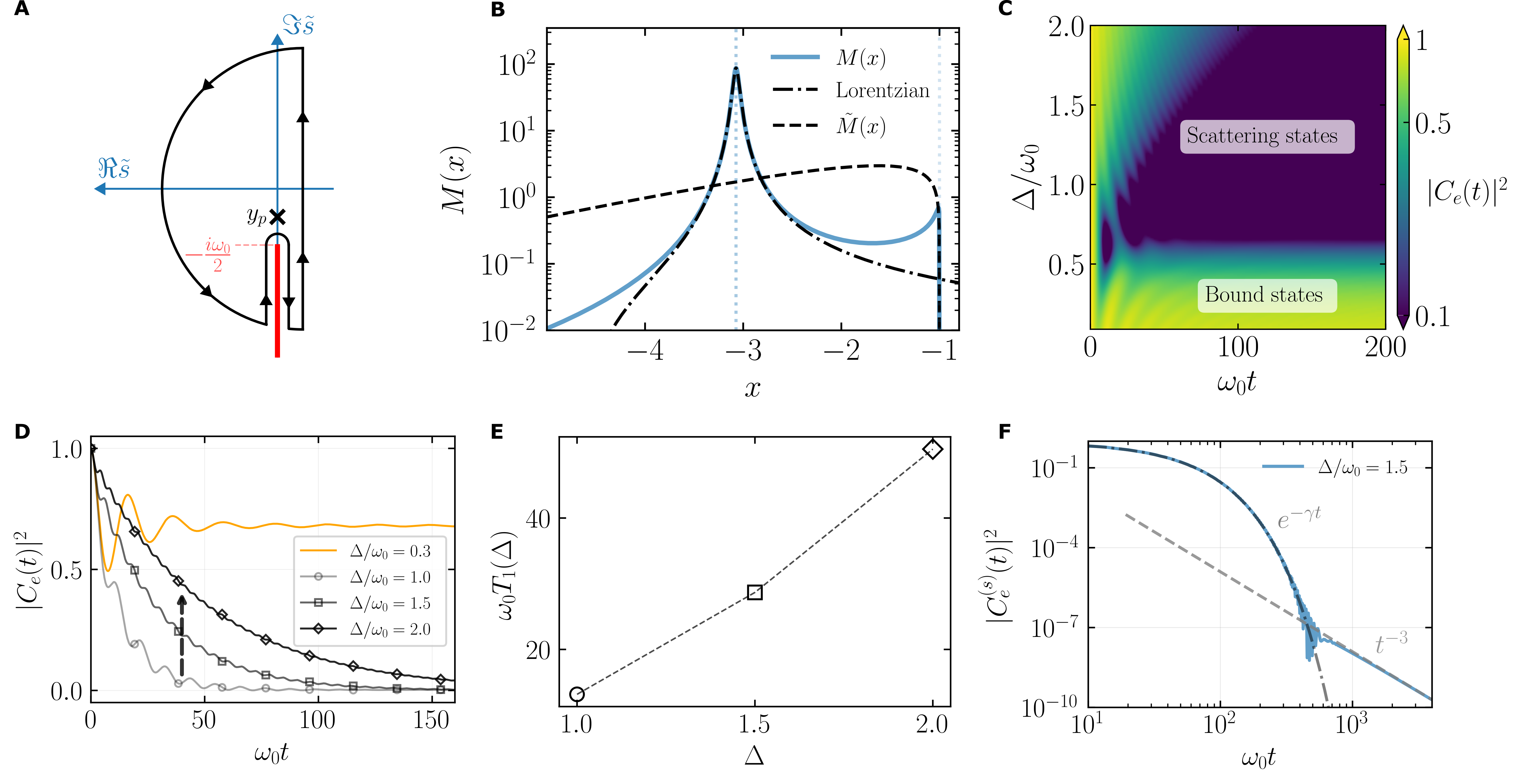}
    \caption{\textbf{Qubit-Dynamics for infinite UV cutoff.}
    \textbf{A} Contour in Laplace $\tilde s$-space used for calculating qubit dynamics. Solid red line and the black cross show the branch cut due to the continuum and bound-state pole $y_{p}$ outside the continuum, respectively.
    \textbf{B} Profile of spectral weight function $M(x)$ plotted in rescaled frequency units $x$ ($=2\omega/\omega_0$), calculated for $\Delta/\omega_0 = 1.5,\, g/\omega_0 = 0.2$. Solid line shows the full function, while patterned lines show the approximations used for capturing the contribution of peak near the dressed emitter frequency ($x \leq 2 \Delta/\omega_0$) (dot-dashed) and contributions from the band-edge (dashed line), respectively.
    \textbf{C} Emitter population dynamics for $\Delta$ across the LHTL band-edge ($\omega_0/2$).
    \textbf{D} Time-slices at different emitter frequencies taken from (C). The arrow indicates increasing relaxation time with increasing $\Delta$ (decreasing decay rate).
    \textbf{E} Relaxation time $T_1 \equiv 1/\omega_0\gamma$ analytically calculated at emitter frequencies in (D).
    \textbf{F} Crossover from exponential relaxation $e^{-\gamma t}$ to long-time power-law tail $t^{-3}$.}
    \label{fig: Qubit-Dynamics}
\end{figure*}
Unlike coupled-cavity arrays and RHTLs which act as Markovian baths to emitter parked sufficiently away from the band edges, the Brownian spectral density for LHTL (Eq.~\ref{eq: Spectral function LHTL flat gk}) leads to highly non-Markovian dynamics even when the emitters are parked deep in-band.  The time-non-local equation for qubit population can be solved exactly using analytical continuation in Laplace domain. Here we report the final form for time-domain solution, assuming infinite cutoff $\omega_{c} \to \infty$ (see Supplementary Materials for details)
\begin{equation}\label{eq: Qubit dynamics C_e(t) integral}
    \begin{split}
        C_e(t) &= \frac{8g^2}{\pi \omega_0^2}\int_{-\infty}^{-1}\,dx\,M(x)\,e^{i\omega_0 x t/2} + r\,e^{iy_pt}\\
        &\equiv C^{(s)}_e(t) + C^{(b)}_e(t).
    \end{split}
\end{equation}
Here $C^{(b,s)}_e(t)$ correspond to the bound-state and scattering contributions respectively. The pole denoted by $y_{p}$ is obtained as a real-valued solution to the bound-state energy equation, ${y + \Delta + \Sigma(-y) = 0}$, with the corresponding residue, $r = [1-\Sigma'(-y_p)]^{-1}$. As evident from the contour shown in Fig.~\ref{fig: Qubit-Dynamics}A, the dominant contribution to $C^{(s)}_e(t)$ coming from the integration about the `continuum branch cut' shown in red.  We refer the function $M(x)$, plotted in Fig.~\ref{fig: Qubit-Dynamics}B, as the spectral weight function which shows a peak near the dressed qubit frequency. Since $C^{(s)}_e(\infty) =0$, the steady-state solution is entirely governed by the contribution from the bound state, $|C_e(\infty)|^2 = |r|^2$.
\par
Finite-time population dynamics are a combination of both scattering and bound-state contributions, though the functional form of decay envelope is dominated by the scattering contribution $C^{(s)}_e(t)$ as different time scales selectively probe different regions of the spectral weight function $M(x)$ (see Fig.~\ref{fig: Qubit-Dynamics}C). We analyze the scattering contribution $C^{(s)}_e(t)$ in three limiting cases. \newline
(i) At very short times, $\max_{k}|\omega(k)-\Delta|\,t\ll 1$, we obtain universal quadratic decay associated with Zeno physics,
\begin{equation}
|C^{(s)}_e(t)|^2 \approx 1-(\Gamma t)^2,
\end{equation}
where $\Gamma={(2\pi)^{-1}}\!\int d\omega\,J(\omega)$. This ``Zeno window" always exists, but its span shrinks as the $\omega_{\rm UV}$ increases. \newline 
(ii) In the infinite-cutoff limit, the decay begins with a Wigner-Weisskopf-type exponential for $0\leq t\lesssim \tau_1\equiv 1/\gamma$. In this regime, the qubit primarily samples modes near the renormalized (ore dressed) qubit frequency, where $M(x)$ can be well approximated by a Lorentzian (see Fig.~\ref{fig: Qubit-Dynamics}D). This yields
\begin{equation}\label{eq: Intermediate time qubit population, exponential decay}
    |C^{(s)}_e(t)|^2 \approx \left(\frac{16 g^2 |a_0|}{\omega_0^2}\right)^2 e^{-\gamma t},
\end{equation}
with $\gamma>0$ determined by $\Delta$, and $a_0\in\mathbb{C}$ is an overall complex detuning-dependent prefactor (see Supplementary Materials). Interestingly, qubit decay rate $\gamma$ decreases with increasing qubit frequency (see black arrow), a direct consequence of inverted spectral density $(J(\omega) \sim 1/\omega^2)$ of the LHTL (Fig.~\ref{fig: Qubit-Dynamics}E).  \newline
(iii) At late times, $t \gg \tau_2$, the region near $x = -1$ \cite{Dynamical_signatures_2017} becomes important where $M(x)$ shows a sharp discontinuity, since rapid oscillations suppress all contributions except those from the lower band-edge. Approximating $\tilde M(x)\simeq M(x \approx -1)$ in this regime gives the long $t^{-3}$ tail, as shown in Fig.~\ref{fig: Qubit-Dynamics}F (see Supplementary Materials), 
\begin{equation}\label{eq: long-time qubit polynomial asymptote}
    |C^{(s)}_e(t)|^2 \approx \frac{1}{16\pi}\left(\frac{\omega_0}{g}\right)^4 \frac{1}{(\omega_0 t)^3}.
\end{equation}
As noted in \cite{Dynamical_signatures_2017}, there is also an intermediate window $\tau_1\,<\,t\,<\,\tau_2$  when the population decays as $t^{-1}$, arising from the contribution of the smaller shoulder near the band edge at $x = -1$) which becomes increasingly sharp as $g\to 0$. If on the other hand, the qubit is parked outside the continuum, see Fig.~\ref{fig: Qubit-Dynamics}, a bound-state is formed  leading to fractional decay of population at long times.
%
\subsubsection*{Photon Dynamics: accelerated light cones}
\label{subsec: Photon Dynamics: accelerated light cones}
%
%
\begin{figure*}[!t]
    \centering
    \includegraphics[width = 0.75\linewidth]{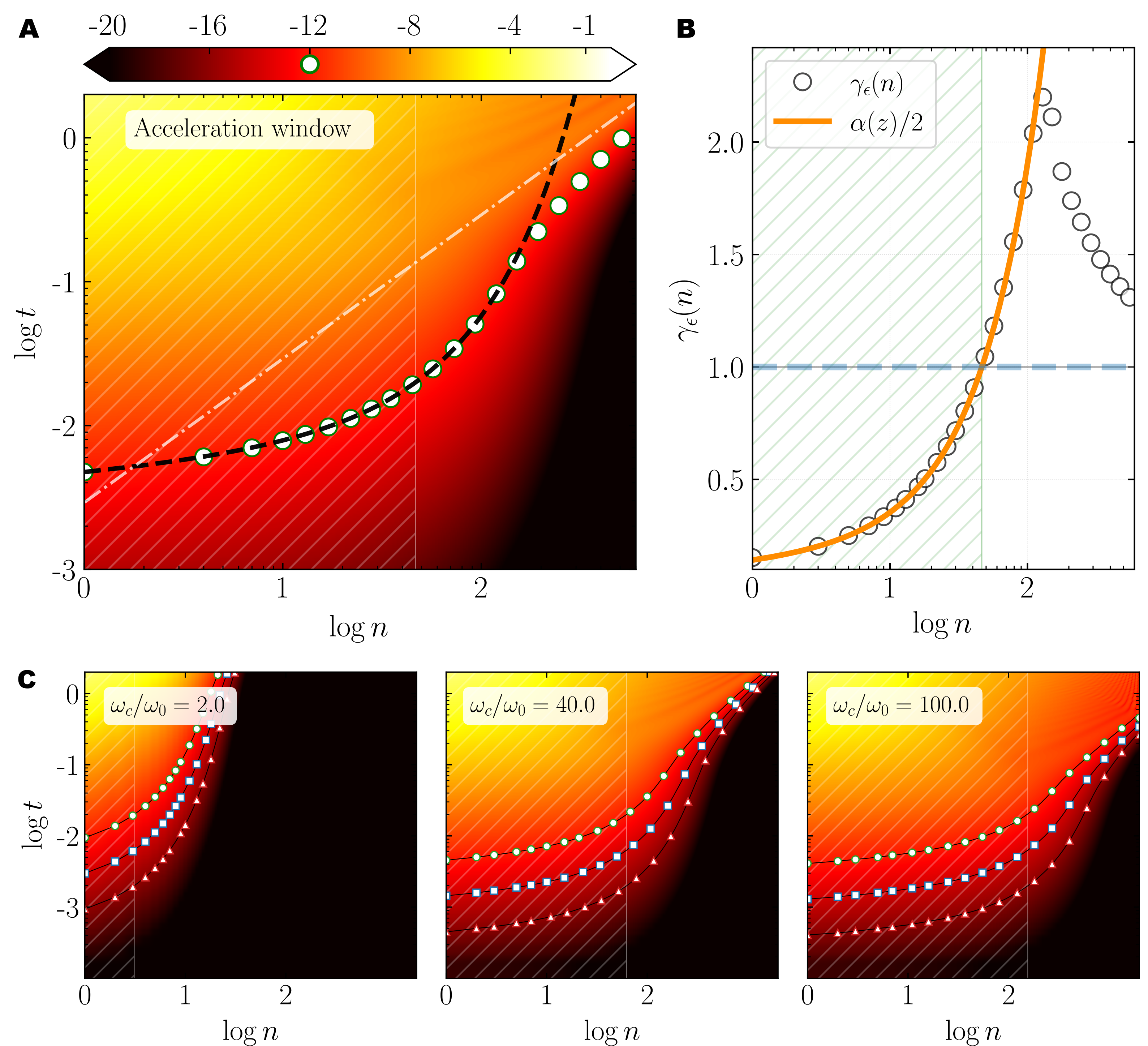}
    \caption{\textbf{Accelerated light cones in LH waveguide-QED.} \textbf{A} Simulated space-time diagrams of photon intensity $\ln |C_n(t)|^2$ for a qubit parked inside the pass band of LHTL ($\Delta/\omega_0 = 1$, $\omega_c/\omega_0 = 30$, $g/\omega_0 = 0.08$). Simulated  constant intensity fronts, $t_\epsilon(n)$ for $\epsilon = 10^{-12}$, are plotted as solid white-circles $(N = 1501)$, while the short-time analytical result is shown with the dashed-black line. Dot-dashed white line corresponds to the linear light cone $t = n/v_\text{max}$. \textbf{B} Empty circles represent the local slope for light-cone obtained from simulations, and the solid-orange curve represents the analytical result obtained in Eq.~\ref{eq: early-time local cone exponent}. \textbf{C} show the space-time density profiles of scattered photons within the accelerated windows for three different choices of $n_{\star} = 2,\, 40,\, 100$. In each density plot, circle, square, and triangle markers show $\epsilon$-fronts for three different choices of $\log(\epsilon) = -12$, $-14$, and $-16$, respectively ($N = 4001$).}
    \label{fig: Photon Dynamics and light-cones}
\end{figure*}
Long-range couplings and interactions are known to profoundly modify the propagation of excitations and correlations in many-body systems \cite{Tran2021LiebRobinson,Cheneau2012-qv,Jurcevic2014-vh,Richerme2014-qz}. 
Specifically, the long-range lattice models can exhibit a gamut of distinct propagation phenomena, ranging from non-local spreading of correlations in strong long-range models to quasi-local propagation (nonlinear light cones). Given the challenges imposed by nonlocal nature of such models, only recently, the optimal bounds on interaction exponents, $\alpha > 2(3)$ in non-interacting (interacting), required for linear light cones in long range models have been reported \cite{Tran2020Hierarchy}. Since LHTL realizes a `running' hopping exponent $\alpha(z)$, this makes it a unique system to access all these regimes in a single setup probed at different length scales.
\par
\begin{table*}[t!]
    \centering
    
    \renewcommand{\arraystretch}{1.2}
    \setlength{\tabcolsep}{6pt}

    \newcommand{\tallrow}{\rule[-2.0ex]{0pt}{6.0ex}}

    \begin{tabular}{c|c|cc|cc|cc}
        \hline\hline

        &
        \multicolumn{1}{c|}{Hopping}
        & \multicolumn{2}{c|}{Bound photon $\beta(z)$}
        & \multicolumn{2}{c|}{Scattered photon $\gamma(z)$}
        & \multicolumn{2}{c}{Spectral function $J(\omega)$}
        \\
        \cline{3-8}
        \noalign{\vskip -0.1ex}
        &
        $\alpha(z)$
        & $g$ & $g_k$
        & $g$ & $g_k$
        & $g$ &  $g_k$
        \\ \noalign{\vskip -0.2ex}
        \hline

        LHTL \tallrow
        & $z\,\dfrac{K_1(z)}{K_0(z)}$
        & $2+\dfrac{z^2}{\alpha(z)}$
        & $\dfrac{1}{2}+z\dfrac{K_{3/4}(z)}{K_{1/4}(z)}$
        & $\dfrac{\alpha(z)}{2}$
        & $z\dfrac{K_{3/4}(z)}{K_{1/4}(z)}$
        & $\sim \dfrac{1}{\omega^2}$
        & $\sim \omega$
        \\

        \hline

        RHTL \tallrow
        & $2$
        & $2$
        & $\dfrac{3}{2}$
        & $1$
        & $\dfrac{3}{2}$
        & $\sim \omega^0$
        & $\sim \omega$
        \\
        \hline

        \multicolumn{1}{c|}{%
            \begin{tabular}{@{}c@{}}
                NN \vspace{-1mm}\\
                $(n_\star<1)$
            \end{tabular}
        } \tallrow
        & $z$
        & $2z$
        & $2z$
        & $1$
        & $1$
        & $\sim \omega^0$
        & $\sim \omega^3$
        \\

        \hline\hline
    \end{tabular}

    \caption{Summary of scaling exponents and spectral density for different TL-based waveguide QED systems. Constant values in exponents correspond to power-law decays while scaling with position $(z)$ indicates exponential falloffs. In the last row, we report the coupled-cavity-array (CCA) regime, characterized by nearest-neighbor photonic hopping, recovered from the LHTL in the limit $n_\star < 1$.}
    \label{tab:exactg_k}
\end{table*}
Here, we adopt a different route to analyze photon propagation in LHTL in order to maintain a thematic connection with position-dependent hopping network derived earlier, which is distinct from the standard methods used in the context of non-interacting bosonic systems such as state transfer protocols or quantum walks on $d$-dimensional lattices \cite{Tran2020Hierarchy}. We analyze photon propagation in LHTL waveguide-QED systems by tracking constant-intensity photonic fronts which we henceforth allude to as $\epsilon$-fronts. We identify $t_\epsilon(n)$ as the earliest time an intensity threshold $\epsilon$ is reached at a distance $n$ along the transmission line. In other words, $t_\epsilon(n)$ tracks the iso-intensity curve in $(n,t)$-plane, i.e., $|C_n(t)| = \epsilon$. For very short-times this yields a direct relationship between the emitted field and the hopping network of the transmission line, $C_n(t) \simeq -\frac{ig}{2} \xi_n t^2$ (Fig.~\ref{fig: Photon Dynamics and light-cones}A). This can be inverted to obtain 
\begin{equation}
t_\epsilon(n) = \sqrt{2\epsilon/g} \, {\xi_n}^{-1/2},
\end{equation}
resulting in a `running' light-cone exponent as
\begin{equation}
\gamma_\epsilon^{\rm LH}(z)\equiv\frac{d\ln t_\epsilon(z)}{d\ln z} = \frac{\alpha^{\rm LH}(z)}{2}.
\label{eq: early-time local cone exponent}
\end{equation}
In 1D free long-range hopping models with constant $1/r^\alpha$, the light cone is rigorously shown to be linear when $\alpha > d+1 = 2$, and no linear cone exists for $1 < \alpha < 2$ \cite{Tran2020Hierarchy}. For $\alpha < 1$, the dynamics are non-local with the details of propagation highly dependent on the choice of model. However, the LHTL hopping network is not constant-$\alpha$–its exponent $\alpha(z)$ grows with distance and crosses $\alpha = 1,2$ at $z_{1} \approx 0.60$, and $z_2 \approx 1.55$ (Fig.~\ref{fig: Photon Dynamics and light-cones}B). Accelerated fronts appear in the region $z<z_2$, where $\alpha(z) < 2$ or $\gamma < 1$. Once $\alpha(z) > 2$, $\gamma \to 1$, the fronts cease to accelerate and align with the linear cone set by $v_{\rm max} = \max_{k} v_g(k)$ at large distances. In fact, in simulation we find the $\epsilon$-fronts decelerate for $z>z_2$ saturating to linear cones in the long distance regime. Unsurprisingly, while the crossover into decelerating light cones is captured by short-time analytics, the long distance saturation to $\gamma = 1$ requires going beyond the linear approximation to retarded photon propagator (see Supplementary Materials). 
\par
This analysis shows that the free light cone bounds derived for systems with constant-$\alpha$ apply locally to the scale-dependent light-cone exponent $\gamma(z)$, which changes continuously as the photon propagates through an LHTL. As all distances $z$ are defined w.r.t. $n_\star = \omega_{\rm UV}/(2 \omega_{\rm IR})$, as in the case of bound state localization, the interplay of UV and IR cutoffs controls the distance over which the accelerated light cones are realized (Fig.~\ref{fig: Photon Dynamics and light-cones}C). Within the accelerating window ($z<z_2$), Eqs.~\ref{eq: BS envelope local exponent} and \ref{eq: early-time local cone exponent} can be combined to obtain a direct relation between the bound-state and light-cone exponents,
\begin{equation}\label{eq: flat g bound-state and light-cone exponent}
    \beta^\text{LH}(z) = 2 + \frac{z^2}{2\,\gamma^\text{LH}(z)}.
\end{equation}
\par
Since, Eq.~(\ref{eq: early-time local cone exponent}) is independent of the handedness of the transmission line, we can use it to study the light cone exponents in RHTLs too. Since  $\alpha^\text{RH}(z) = \bar{\alpha}^\text{RH}(z) = 2$ (Eq.~\ref{eq:alpha-RH}), the $\epsilon$-cones are linear, $\gamma^\text{RH}(z) = \alpha^\text{RH}(z)/2 = 1$, at all times. Hence, in contrast to LHTL, $z_2^{\rm RH} = 0$ implying that there is no region in an RHTL waveguide-QED system where the free light cones are super-linear or accelerated (see Supplementary Materials). This is in alignment with the extended Lieb-Robinson bounds for systems with power-law hopping \cite{Tran2020Hierarchy} which predict that a finite-sized acceleration window ($\gamma<1$) requires hopping that decays \emph{slower} than at least $1/n^2$ ($\alpha < 2$) over that distance. The RHTL case with $\xi_n\,\sim\,1/n^2$ falls in the marginal, non-accelerating limit.
%
\subsection*{Robustness of the long-range physics to coupling structure} 
\label{sec: Effects of coupling structure}
Most theoretical treatments of standard waveguide-QED consider spatially local emitter-waveguide coupling, while ignoring its resultant momentum-dependence. This assumption is strictly true only for weak coupling to environments with a small dispersion bandwidths. Even in the presence of structured environments, specifically which lead to a non-white or `colored' $J(\omega)$, such an approximation maybe more forgiving for calculating emission properties, such as relaxation rates, where the density of states near resonance with the emitter dictate the dynamics. Over such narrow spectral windows, one can always make a linear expansion of dispersion near the relevant frequency leading to a `white' $J(\omega)$.  The validity of such an approach, however, remains in question when calculating properties of bound states that are determined by the global structure (`poles') of the self-energy function. Moreover, such local approximations may fail even for scattering properties when propagation over extended distances is relevant, such as, for calculation of light cone exponents or relaxation of atoms in situations which fall outside the purview of standard dipole approximation (e.g., giant atoms\cite{Andersson2019}, phononic baths etc.). The structural non-Markovianity and native long-range nature of transmission line-based waveguide-QED systems, makes it seem more imperative to revisit the assumption of assuming of flat $g_{k}$ for both RHTL and LHTL cases. From a detailed microscopic derivation, we find $g_{k}^{\rm LH} \sim \omega_{k} ^{3/2}$ and $g_{k}^{\rm RH}(k) \sim \omega_{k}^{1/2}$ (See Supplementary Materials).

\begin{figure}[t!]
    \centering
    \includegraphics[width=0.85\linewidth]{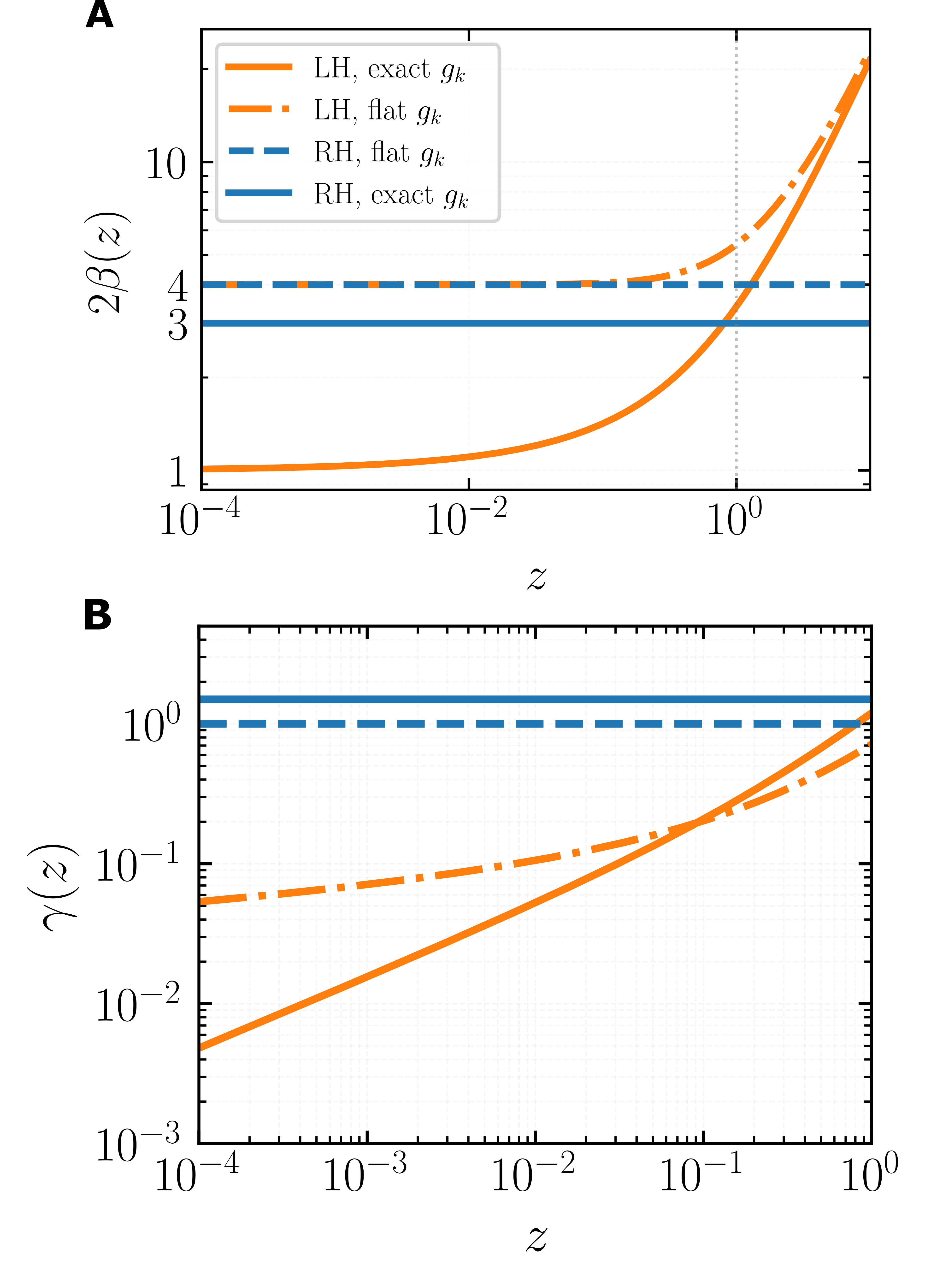}
    \caption{\textbf{Effect of momentum-dependent emitter-waveguide coupling:} \textbf{A} Bound-state exponent $2\beta(z)$ for flat (dashed) vs momentum-dependent (solid) coupling constant. The long-range nature of LHTL manifests in position-dependent exponents obtained for both cases (orange), while RHTL shows constant-exponent power-law behavior in both cases (blue). \textbf{B} Light-cone exponents $\gamma(z)$ calculated for short-time regime. Faster acceleration is obtained with $g(k)$, as compared to constant coupling, for LHTL (orange) while the reverse is true for RHTL (blue). Interestingly, there is a crossover at $z \approx 0.1$, which implies that over distances in the range $0.1 < z <z_{2}$ the flat g cones are more accelerated than their momentum-dependent counterparts. Such a crossover is not present in case of localization exponents.}
    \label{fig: Effects of microscopic coupling}
\end{figure}

Table \ref{tab:exactg_k} reports the effect of momentum-dependence of $g_{k}$ and quantify its effect on both bound state $\beta(z)$ and light cone exponents $\gamma(z)$ for both right- nad left-handed transmission line systems. One can presage the expected modification here by noting that while the couplings increase with frequency in both cases, they can have distinct and even opposing effects on extended spatial behavior due to opposite nature of dispersion in the two TL systems: while in LHTL this boosts the participation of long-wavelength modes, in RHTL it privileges short-wavelength contributions instead. Interestingly, compared to Eq.~\ref{eq: BS envelope local exponent}, including the momentum-dependence of $g(k)$ leads to a significantly lower local bound-state exponent $\beta(z)$ for $z\ll1$ (Fig.~\ref{fig: Effects of microscopic coupling}A). This implies an even weaker bound-state localization with LHTL waveguide, that exhibits a spatial rolloff as $1/n$ instead instead of $1/n^{4}$ , which can be attributed to an enhanced participation of high-frequency or large-wavelength modes. On the other hand, RHTL experiences a relatively minor dilution of $\beta(z)$ to $3$ ($\forall z$) as compared to that obtained for flat coupling profiles. 
\par
Similar treatment for LHTL light cone exponents shows enhanced acceleration, as reflected in numerically smaller values of $\gamma(z)$ compared to those obtained for the flat $g$ (Fig.~\ref{fig: Effects of microscopic coupling}B). For RHTL case, the light cones are decelerated at all distances (up to valid short-time regime), as ${\gamma_\text{RH} (z) = 3/2 > 1}$, due to momentum dependence of couplings. Surprisingly, momentum-dependent coupling profile leads to a simpler relation between the bound-state and light-cone exponents, 
\begin{equation}\label{eq: exact gk bound-state and light-cone exponent}
    \beta^\text{LH}(z) = \frac{1}{2} + \gamma^\text{LH}(z),
\end{equation}
as compared to the expression in Eq.~\ref{eq: flat g bound-state and light-cone exponent} obtained for momentum-independent flat $g$; this perhaps can be attributed to even more dominant participation of long-wavelength modes which simplifies the position-dependence of coefficients entering the relationship between exponents.
%
\section*{Discussion}
%
In summary, we showed that transmission line waveguide-QED systems emulate \emph{native} long-range photonic hopping networks that can support new bound-state localization phenomena. While RHTL maps to a tight-binding lattice with hoppings that decay quadratically with distance, LHTL systems in contrast show a logarithmic decay of hoppings upto a length scale $n_\star = \omega_\text{UV}/2\omega_\text{IR}$. 
\par
It is worth noting that in most waveguide-QED studies, photon localization and photon propagation are treated as independent phenomena. Here, using the running-exponent method we derive analytical relationships between the bound state local exponent extracted from spatial profile of localized wavefunction ($\beta(z)$) and light cone exponent extracted from light fronts of scattered photons ($\gamma (z)$) respectively, both for flat (momentum-independent) and structured (momentum-dependent) emitter-waveguide coupling (Eqs.~\ref{eq: flat g bound-state and light-cone exponent}, and \ref{eq: exact gk bound-state and light-cone exponent}). Crucially, the crossover from algebraic to exponential bound-state localization as well as from accelerated to linear propagation fronts is set by the scale $n_\star$, making it a unique knob to tune interaction range in LHTL waveguide-QED systems. 
\par
Employing the language of running exponents for analyzing TL-based waveguide-QED also allows us to contrast the distinctions between constant-exponent (RHTL) and running-exponent (LHTL) hopping networks. Interestingly, this clarifies that a simultaneous presence of both these phenomena is predicated on realizing strong long-range couplings, a regime inaccessible in RHTL; thus RHTL waveguide-QED exhibits only linear light cones even though it realizes scale-invariant power-law localization. Our results are in direct agreement with the free light cone bounds derived for non-interacting power-law systems, using a single-particle state transfer picture \cite{Tran2020Hierarchy}. More generally, this presents a fresh perspective on engineering soft modes and non-analytical features in dispersion of low-dimensional many-body systems, such as the pole at $k=0$ in LHTL, which have previously been identified as the harbinger of asymptotic long-range interactions \cite{LimitsofPhoton-mediated2020}.
\par
Our findings constitute a paradigmatic change in waveguide-QED and offer several immediate extensions: most foundational results in single-excitation regime are derived for short-range waveguide models, such as coupled-cavity arrays, and need to be revisited for intrinsically long-range waveguides. Further, structural non-Markovianity and accelerated photon scattering (e.g. super-radiance and sub-radiance) in such systems can lead to strong modifications of multi-photon cooperative effects such as super- and sub-radiance, and time-delayed feedback control \cite{Grimsmo2015}. Multi-emitter LHTL systems with tunable power-law exchange interactions can lead to new capabilities for analog quantum simulations \cite{Buluta2009,Zhang_science2023} of many-body models (eg. extended Bose–Hubbard, long-range spin chains), and help explore new dynamical regimes in engineered open quantum systems and unconventional spin-boson models. 
\par
From an application perspective, since superconducting LHTLs have already been experimentally demonstrated \cite{WangLHTL2019}, the proposed platform is within reach of current circuit-QED capabilities and qualitative signatures identified here accessible to experimental tests in the near future. Beyond serving as unconventional waveguides, nonlinear LHTLs embedded with Josepshon junctions have already attracted growing interest as a new class of traveling-wave parametric amplifiers due to their exceptional gain characteristics \cite{Kow2025}. Additionally, topological edge states in nonlinear LHTLs have also been exploited for efficient higher harmonic generation \cite{Wang2019}. This suggests a broad design space available to microwave waveguide-QED systems via a combination of dispersion, nonlinearity and topological properties. Leveraging such platforms to engineer hardware-native nonlocal couplings between distant qubits can also serve to alleviate current bottlenecks in wiring and connectivity of distributed quantum networks, enabling new resource-efficient multi-qubit architectures for long-range entanglement distribution \cite{Singh2025}. 
%
%
\section*{Methods}
%
\emph{Running exponent method}--- Any positive, monotonically decreasing function $\xi(z)$ can be parameterized using \newline
(i) running local exponent

\begin{equation}
\alpha(z)\;\equiv\;-\frac{d\ln \xi(z)}{d\ln z},
\label{eq:alpha-def}
\end{equation}

with monotonicity implying $\alpha(z) > 0$,
and \newline
(ii) running global exponent $\bar{\alpha}(z;z_0)$
\begin{equation}
\begin{split}
    \xi(z)&=\xi(z_0)\Big(\frac{z}{z_0}\Big)^{-\bar{\alpha}(z;z_0)},\\
\bar{\alpha}(z;z_0)&=\frac{1}{\ln(z/z_0)}\int_{z_0}^{z}\frac{\alpha(z')}{z'}\,dz'
\end{split}
\label{eq:alpha-bar-def}
\end{equation}
The location $z_0 = 1/n_\star$ is the unit-cell adjacent to the location of the emitter. We use this method to identify the regime for strong vs weak-long range as well as for explaining spatial profiles of bound states and temporal profiles of constant-intensity light fronts for scattering states presented in this work.  
%
%
%

%
\section*{Acknowledgments}
%
The authors are grateful to David Zueco for a careful reading of the manuscript and helpful pointers about the presentation. P.G. would like to acknowledge Saptarshi Biswas for insightful discussions on long-range models. 
\paragraph*{Funding:}
This research was supported by National Science Foundation under grant
number DMR-2508447 (P.G.), Defense Advanced Research
Projects Agency (DARPA) Synthetic Quantum Nanostructures (SynQuaNon) program under grant number HR00112420343 (J.L.), and Department of Energy under grant DE-SC0019461 (C.A.G-G.). 
 C.A.G-G acknowledges financial support from SECIHTI Mexico, under
grant number CBF2023-2024-2888, and DGAPA-PAPIIT-UNAM under grant number IA104625.
\paragraph*{Author contributions:}
A.K. and C.A.G.G. conceptualized the project. P.G. developed the analytical framework under the supervision of A.K., aided by initial calculations performed by C.A.G.G. J.L. performed the numerical diagonalization methods used here for benchmarking analytical predictions. A.K. and P.G. wrote the manuscript in discussion with other authors.
\paragraph*{Competing interests:} There are no competing interests to declare.
\paragraph*{Data and materials availability:} All data needed to evaluate the conclusions in the paper are present in the paper and/or the Supplementary Materials. Additional data related to this paper may be requested from the authors.
%
\subsection*{Supplementary Materials}
\noindent
S1. Circuit Quantization for left- and right-handed transmission lines\\
S2. Self-energy: Decay Rate and Lamb Shift\\
S3. Algebraic Bound-state\\
S4. Bound-state core: waveguide-QED near the band-edge\\
S5. Reproducing the dispersion by extended hopping\\
S6. Exact Qubit Dynamics\\
S7. Light-cone expression\\
S8. Numerical Diagonalization\\
Figure S1. Right-handed waveguide-QED.\\
Figure S2. Self-energy function.\\
Figure S3. Numerical validation of bound-states.\\
Figure S4. LHTL Bound-state core exponential window.\\
Figure S5. Dispersion relation.\\
Figure S6. Light-cones in LHTL and RHTL.\\

%
\end{document}



\renewcommand{\thefigure}{S\arabic{figure}}
\renewcommand{\thetable}{S\arabic{table}}
\renewcommand{\theequation}{S\arabic{equation}}
\renewcommand{\thesubsection}{S\arabic{section}.\arabic{subsection}}
\renewcommand{\thesection}{S\arabic{section}}
\setcounter{figure}{0}
\setcounter{table}{0}
\setcounter{equation}{0}
\setcounter{page}{1}

\author{
	P. Goswami$^{1}$,
	J. Liu$^{1}$, C. A. Gonz\'{a}lez-Guti\'{e}rrez$^{2,3}$, 
    A. Kamal$^{1,2}$\and
	\small$^{1}$Department of Physics and Astronomy, Northwestern University, Evanston, IL 60208, USA.\and
	\small$^{2}$Department of Physics and Applied Physics, University of Massachusetts, Lowell, MA 01854, USA.\and
    \small$^{3}$Instituto de Ciencias F\'{i}sicas, Universidad Nacional Aut\'{o}noma de M\'{e}xico, Cuernavaca 62210, Mexico \and
	\small$^\ast$A. Kamal. Email: archana.kamal@northwestern.edu
}

\newcommand{\PaperTitle}{\scititle}
\newcommand{\PaperAuthors}{P. Goswami$^{1}$, J. Liu$^{1}$, C. A. Gonz\'{a}lez-Guti\'{e}rrez$^{2,3}$, A. Kamal$^{1,2}$}
\newcommand{\PaperAffils}{%
\small$^{1}$Department of Physics and Astronomy, Northwestern University, Evanston, IL 60208, USA.\\
\small$^{2}$Department of Physics and Applied Physics, University of Massachusetts, Lowell, MA 01854, USA.\\
\small$^{3}$Instituto de Ciencias F\'{i}sicas, Universidad Nacional Aut\'{o}noma de M\'{e}xico, Cuernavaca 62210, Mexico\\
\small$^\ast$A. Kamal. Email: archana.kamal@northwestern.edu
}

\begin{center}
{\Large \textbf {Supplementary Materials for\\ \PaperTitle}\par}
\vspace{0.5em}
{\PaperAuthors\par}
\vspace{0.25em}
{\PaperAffils\par}
\end{center}

\subsubsection*{This PDF file includes:}
\noindent
S1. Circuit Quantization for left- and right-handed transmission lines\\
S2. Self-energy: Decay Rate and Lamb Shift\\
S3. Algebraic Bound-state\\
S4. Bound-state core: waveguide-QED near the band-edge\\
S5. Reproducing the dispersion by extended hopping\\
S6. Exact Qubit Dynamics\\
S7. Light-cone expression\\
S8. Numerical Diagonalization\\
Figure S1. Right-handed waveguide-QED.\\
Figure S2. Self-energy function.\\
Figure S3. Numerical validation of bound-states.\\
Figure S4. LHTL Bound-state core exponential window.\\
Figure S5. Dispersion relation.\\
Figure S6. Light-cones in LHTL and RHTL.\\
\clearpage
%
\section{Circuit Quantization for left- and right-handed transmission lines}
%
In this section, we derive the circuit quantization of the composite transmon–transmission-line system and obtain the corresponding single-excitation waveguide-QED Hamiltonian. We first consider the left-handed waveguide and then generalize the same notation and framework to the right-handed case. Assuming weak capacitive coupling between the transmon and the transmission line, we quantize the uncoupled subsystems separately and incorporate the interaction Hamiltonian afterwards. This is in line with the perturbative RWA Hamiltonian shown in Eq. 2 of the main text. \newline

\subsection{Quantization of a left-handed waveguide}
\subsubsection{LHTL Hamiltonian}
The left-handed transmission line (LHTL), shown in Fig.~1A in the main text, is described by the Lagrangian written in node-flux basis
\begin{equation}
\mathcal{L} = \sum_{n=0}^{N-1}
\left[
\frac{C_0}{2} \left( \dot{\Phi}_{n} - \dot{\Phi}_{n-1} \right)^2
+\frac {C}{2}\dot{\Phi}^2_{n}
- \frac{\Phi_{n}^2}{2L}
\right].
\end{equation}
We impose periodic boundary conditions, $\Phi_{N} = \Phi_{0}$, as boundary effects are not the focus here. Under this assumption, the analysis below becomes exact in the thermodynamic limit used in the main text. We introduce the discrete Fourier transform (DFT) from the site basis $n$ to the momentum basis $k$,
\begin{equation}
\Phi_n
=
\frac{1}{\sqrt{N}}
\sum_k
e^{i n k d}\,
\tilde{\Phi}_k ,\quad k \in \left[-\frac{\pi}{d},\,\frac{\pi}{d}\right)
\end{equation}
where $N$ is the number of flux nodes in the LHTL, $d$ is the unit-cell spacing, and the allowed momenta lie in the first Brillouin zone with spacing $\Delta k = 2\pi/(Nd)$. As we show below, the Lagrangian is diagonal in momentum space and can be written compactly in terms of the flux vector as
\begin{equation}\label{SMeq: Original Lagrangian}
\mathcal{L} = \frac{1}{2}\dot{\bm{\Phi}}^\dagger \bm{V}\dot{\bm{\Phi}} - \frac{1}{2L}\bm{\Phi}^\dagger \bm{\Phi}, \qquad \bm{\Phi} = [\Phi_0,\Phi_1,\ldots,\Phi_{N-1}]^\dagger,
\end{equation}
where the matrix $\bm{V}$ is given by
\begin{equation}
\bm{V} \equiv
C_0\begin{pmatrix}
2+\eta & -1 & & & \cdots & -1 \\
-1 & 2+\eta & -1 & & & \\
& -1 & 2+\eta  & -1 & & \\
& & -1 & 2+\eta  & \ddots & \\
\vdots & & & \ddots & \ddots & -1 \\
-1 & & & & -1 & 2+\eta
\end{pmatrix}_{N \times N}
=
(2C_0+C)\mathbb{I}_N - C_0\bm{T} - C_0\bm{T^\dag}
\end{equation}
with
\begin{equation}
\bm{T} =
\begin{pmatrix}
0 & 1 & & &  0  \\
 & 0 & 1 & &    \\
& &  &  \ddots &   \\
 & & &   0 & 1 \\
1 & & & &0
\end{pmatrix}_{N \times N},
\qquad
\eta = \frac{C}{C_0}.
\end{equation}
The DFT can be expressed in a compact matrix form as $\bm{\Phi} = \bm{U}\tilde{\bm{\Phi}}$, where the unitary matrix is given as $(\bm{U})_{nm} = N^{-1/2} \exp[{-in(\pi - {2\pi m}/{N})}],\;\text{for } m,n \in [0,N-1]$. The Lagrangian in $k$-space then reads, 
\begin{equation}
\mathcal{L}
= \frac{1}{2} \dot{\tilde{\bm\Phi}}^\dagger \bm{U^\dag} \bm{V}\bm{U} \dot{\tilde{\bm\Phi}}
- \frac{1}{2L} \bm{\tilde\Phi}^\dag\, \bm{\tilde\Phi}.
\end{equation}
Using $\bm{U^\dag}\bm{T}\bm{U} = \mathrm{diag}(\{e^{ikd}\})$, we can identify the diagonal matrix,
\begin{equation}
\begin{split}
    \bm{U^\dag} \bm{V}\bm{U}
= \mathrm{diag}(\{C_k\}) \equiv \bm{C}
\end{split}
\end{equation}
where $C_k = 2C_0+C-2C_0\cos(kd)$. Thus the diagonalized Lagrangian in $k$-space thus takes the form
\begin{equation}\label{SMeq: k-space diagonalized vector Lagrangian}
    \mathcal{L} = \frac{1}{2} \dot{\tilde{\bm\Phi}}^\dagger  \bm{C}\, \dot{\tilde{\bm\Phi}}
- \frac{1}{2L} \bm{\tilde\Phi}^\dag \bm{\tilde\Phi}
\end{equation}
The Lagrangian in Eq.~\eqref{SMeq: k-space diagonalized vector Lagrangian} can be expanded as,
\begin{align}
\mathcal{L}
&= \frac 12\sum_k C_k\dot{\tilde\Phi}_k^\dag\dot{\tilde\Phi}_k-\frac 1{2L}\tilde\Phi_k^\dag\tilde\Phi_{k}
=\frac 12\sum_k C_k\dot{\tilde\Phi}_k\dot{\tilde\Phi}_{-k}-\frac{1}{2L}\tilde\Phi_k\tilde\Phi_{-k}.
\end{align}
where in the last step we used the property $\tilde{\Phi}_k^\dagger = \tilde{\Phi}_{-k}$. Using the conjugate charge variables,
\begin{equation}\label{eq: Q_k}
\tilde Q_k \equiv \frac{\partial L}{\partial \dot{\tilde{\Phi}}_k} = C_k\dot{\tilde\Phi}_{-k},
\end{equation}
and Legendre transforming the Lagrangian to the Hamiltonian, we obtain the diagonal form of the LHTL Hamiltonian, 
\begin{equation}\label{SMeq: Classical Hamiltonian in k-space}
 H_\text{LH}
= \sum_k \tilde Q_k\dot{\tilde\Phi}_k- \mathcal{L}
=\sum_k \left(\frac 1{2C_k} \tilde Q_k \tilde Q_{-k}
+ \frac 1{2L}{\tilde\Phi}_k{\tilde\Phi}_{-k}\right).
\end{equation}
%
\subsubsection{Quantization}
%
We define the ladder operators as,
\begin{equation}\label{eq:ladder_ope}
\hat{a}_k \equiv \dfrac{1}{\sqrt{2\hbar}}
\left[
\dfrac{\hat{\tilde\Phi}_k}{\sqrt{Z_k}} + i \sqrt{Z_k} \hat{\tilde Q}_{-k}
\right], \\[10pt]
\hat{a}_k^\dagger \equiv \dfrac{1}{\sqrt{2\hbar}}
\left[
\dfrac{\hat{\tilde\Phi}_{-k}}{\sqrt{Z_k}} - i \sqrt{Z_k} \hat{\tilde Q}_k
\right]
\end{equation}
then it follows
\begin{equation}
\hat{\tilde\Phi}_k = \sqrt{\frac{\hbar Z_k}{2}}
\left( \hat{a}_k + \hat{a}^\dagger_{-k} \right),
\quad
\hat{\tilde Q}_{k} = \frac{1}{i} \sqrt{\frac{\hbar}{2Z_k}}
\left( \hat{a}_{-k} - \hat{a}^\dagger_{k} \right).
\end{equation}
The Hamiltonian in Eq.~(\ref{SMeq: Classical Hamiltonian in k-space}) in terms of ladder operator is given as,
\begin{align}\label{eq:TL_quantization}
\hat{H}_\text{LH}
=\sum_k\hbar\omega_k\hat a_k^\dagger\hat a_k, \qquad \omega_k = \frac{1}{\sqrt{LC_k}}= \frac{\omega_0 \omega_c}{\sqrt{\omega_0^2 + 2\omega_c^2(1-\cos(kd))}}
\end{align}
where $\omega_k$ is the dispersion relation for the LHTL. Imposing commutation relation between creation annihilation operators, 
\begin{equation}
[\hat{a}_{k}, \hat{a}^{\dagger}_{k'}]= \delta_{k,k'}.
\end{equation}
or equivalently,
\begin{equation}
[{\hat{\tilde\Phi}}_k, \hat{\tilde Q}_{k'}]=i\hbar \delta_{k,k'}.
\end{equation}
lead to the following commutation relations for real space charge and flux operators, 
\begin{equation}
    [\hat\Phi_n, \hat Q_m ] = i\hbar\delta_{nm},
\end{equation}
where
\begin{equation}\label{eq:Fourier_Qk}
\begin{gathered}
\hat{\Phi}_n = \frac{1}{\sqrt{N}} \sum_k e^{inkd} {\hat{\tilde \Phi}}_k,\quad \hat{Q}_n = \frac{1}{\sqrt{N}} \sum_k e^{-inkd} {\hat{\tilde Q}}_k.
\end{gathered}
\end{equation}

Note that the DFT convention for charge operator is opposite to that used for transforming the flux operator [see Eq.~(\ref{eq: Q_k})].
%
\subsubsection{Transmon qubit coupled to LHTL}
%
Consider a transmon circuit (qubit) coupled to the transmission line at position $s\in [0,N-1]$. The transmon capacitance is denoted as $C_q$ and the coupling capacitance as $C_c$, the full circuit Lagrangian is written as,
\begin{equation}
\mathcal{L} = \sum_{n=0}^{N-1}\left[ \frac{C_0}{2} \left( \dot{\Phi}_{n} - \dot{\Phi}_{n-1} \right)^2 - \frac{\Phi_{n}^2}{2L}\right] 
\\+\frac 12C_q\dot\phi^2+E_J\cos\left(\frac{\phi}{\varphi_0}\right) 
\\+ \frac 12 C_c\big(\dot\Phi_s-\dot\phi\big)^2 
\end{equation}
where $\phi$ denotes the transmon flux variable. We will use $q$ (qubit) to present transmon charge variables. Considering weakly coupled qubit, $C_c \ll C,\,C_0$, we ensure that it does not change the waveguide dispersion, we approximate coupling term in the Lagrangian using,
\begin{equation}
\begin{aligned}
Q_n &= \frac{\partial L}{\partial \dot{\Phi}_n} = 2C \dot{\Phi}_n - C \dot{\Phi}_{n+1} - C \dot{\Phi}_{n-1} + C_c (\dot{\Phi}_n - \dot{\phi}) \delta_{ns} \approx 2C \dot{\Phi}_n - C \dot{\Phi}_{n+1} - C \dot{\Phi}_{n-1} \\
q &= \frac{\partial L}{\partial \dot{\phi}} = (C_q + C_c) \dot{\phi} - C_c \dot{\Phi}_s \approx (C_q + C_c) \dot{\phi} \approx C_q \dot{\phi},
\end{aligned}
\end{equation}
allowing us to quantize qubit and LHTL Hamiltonian separately. This leads to the full Hamiltonian,
\begin{equation}
\begin{aligned}
H &= \underbrace{\sum_k \left(\frac{1}{2C_k}\hat{\tilde Q}_k \hat{\tilde Q}_{-k} + \frac 1{2L}\hat{\tilde\Phi}_k\hat{\tilde\Phi}_{-k}\right)}_{\hat{H}_\text{LH}} + \underbrace{
\frac{1}{2C_q}\hat q^2-E_J\cos\left(\frac{\hat \phi}{\varphi_0}\right)}_{\hat{H}_q}
-\underbrace{\frac{C_c}{C_q}\hat q\dot{\hat\Phi}_s}_{\hat{H}_\text{int}}
\end{aligned}
\end{equation}
where $\hat{H}_\text{LH}$ is given in Eq.~\eqref{eq:TL_quantization}. Truncating the transmon to lowest two levels by defining,
\begin{equation}
\hat q = \sqrt{\frac{\hbar}{2Z_q}}\frac{1}{i}(\sigma^--\sigma^+),\quad \hat \phi = \sqrt{\frac{\hbar Z_q}{2}}(\sigma^-+\sigma^+),\quad Z_q = \sqrt{\frac{L_q}{C_q}}.
\end{equation}
with $L_q = \varphi_0^2/E_J$, leads to an effective qubit Hamiltonian, 
\begin{equation}
\hat{H}_q = \frac{1}{2C_q}\hat q^2-E_J\cos\left(\frac{\hat \phi}{\varphi_0}\right) \approx \hbar\Delta\sigma^+\sigma_-
\end{equation}
where $\Delta = 1/\sqrt{C_qL_q}$ is the qubit frequency. For the coupling Hamiltonian, we intentionally keep the form $\dot\Phi_s$ in the Hamiltonian, despite that a standard Hamiltonian should only consist of flux and charge variables. Using
\begin{equation}
\tilde Q_{k} = C_{k} \dot{\Phi}_{-k} = C_k\frac 1{\sqrt{N}}\sum_n e^{inkd}\dot\Phi_n
\end{equation}
we can write
\begin{equation}
\dot\Phi_s = \frac{1}{\sqrt{N}} \sum_k e^{-iskd} \frac{\tilde{Q}_k}{C_k}
\\=\frac{1}{\sqrt{N}} \sum_k e^{-iskd}\frac{1}{C_k} \left(\frac{1}{i}\sqrt{\frac{\hbar}{2Z_k}} \left( \hat{a}_{-k} - \hat{a}^\dagger_{k} \right)\right),
\end{equation}
and simplify the coupling term as,
\begin{equation}
\begin{aligned}
\hat{H}_\text{int} &= -\frac{C_c}{C_q}\hat q\dot{\hat\Phi}_s = \sum_k g_ke^{-iskd}(\sigma^--\sigma^+)\left( \hat{a}_{-k} - \hat{a}^\dagger_{k} \right)\\
&\approx -\sum_k\left( g_ke^{-iskd}\sigma^-\hat{a}^\dagger_{k}+g_ke^{iskd}\sigma^+\hat{a}_{k}  \right)
\end{aligned}
\label{eq:LHTLrwa}
\end{equation}
with the momentum-dependent coupling $g_k$ given as~\cite{CarrollRingResonators2024},
\begin{equation}
g_k = g_{0} \left(\frac{C_0}{C}\right)^{1/2}\left(\frac{\omega_k}{\omega_{0}}\right)^{3/2}, \qquad g_0 = \frac{C_{c}}{2\sqrt{N}} \sqrt{\frac{\Delta\omega_{0}}{C_{q} C_{0}}}. 
\end{equation}
Note that in the last step of deriving Eq. (\ref{eq:LHTLrwa}), we have used RWA and $g_{-k} = g_k$.
This leads to the total Hamiltonian shown in Eq. 2 of the main text, 
\begin{equation}\label{SMeq: left-handed waveguide-QED Hamiltonian}
    H = \Delta \sigma^+ \sigma^- + \sum_k \omega_k\hat a_k^\dagger \hat a_k \, + \sum_k\left( g_ke^{-iskd}\sigma^-\hat{a}^\dagger_{k}+g_ke^{iskd}\sigma^+\hat{a}_{k}  \right).
\end{equation}
\begin{figure}[b!]
    \centering
    \includegraphics[width=0.5\linewidth]{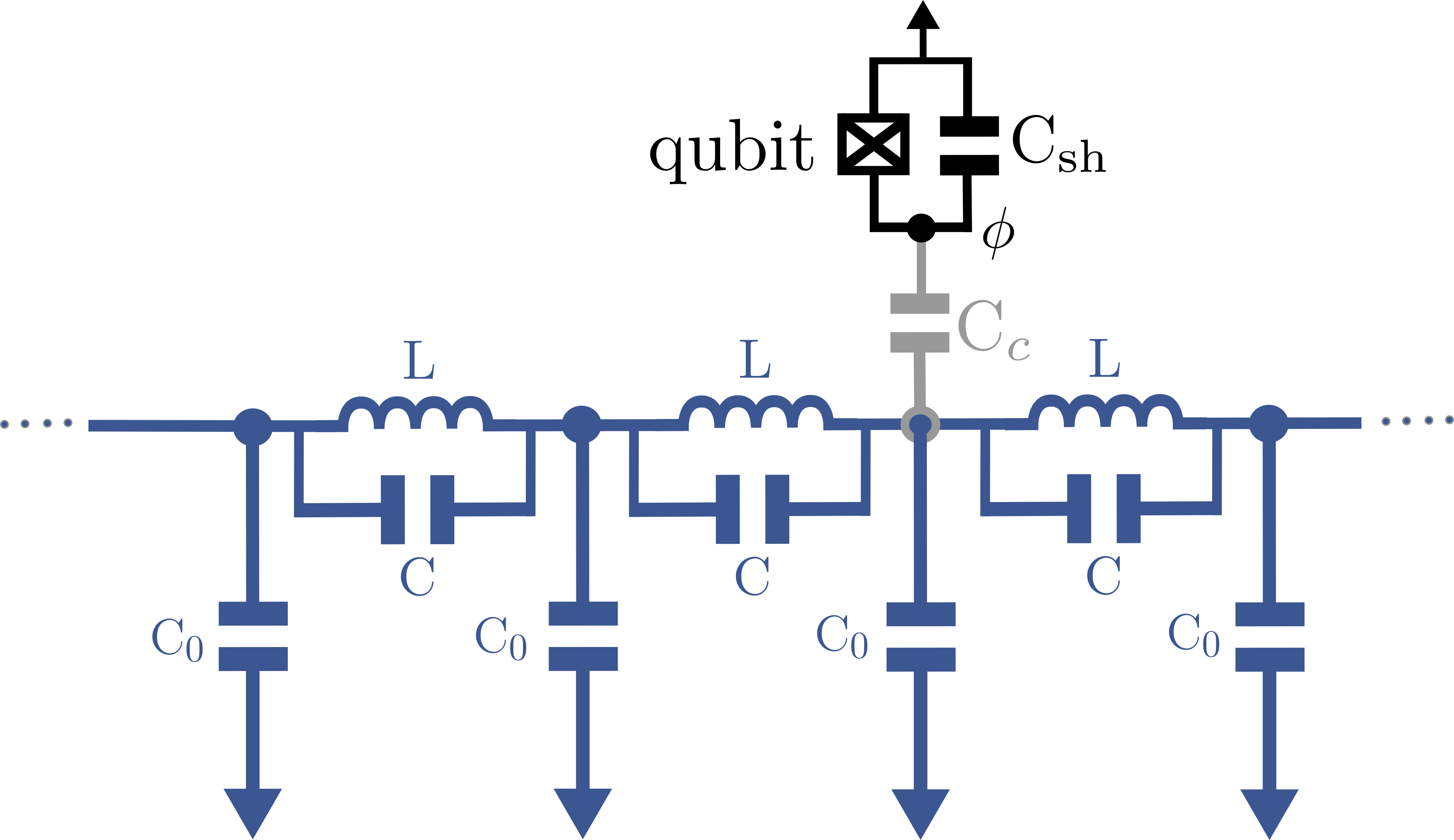}
    \caption{\textbf{Right-handed waveguide-QED.}}
    \label{fig: RH-waveguide}
\end{figure}
\subsection{Quantization of a right-handed waveguide}
For RHTL shown in Fig.~\ref{fig: RH-waveguide}, each unit cell contains an inductor $L$ in series with its associated capacitor $C$ in the main chain, while each node between adjacent LC oscillators is connected to ground via a capacitor $C_0$. The Lagrangian for the circuit is
\begin{equation}
    \mathcal{L} = \sum_{n=0}^{N-1} \frac{C_0}{2}\dot\Phi_{n}^2 +\frac{C}{2}(\dot\Phi_{n}-\dot\Phi_{n-1})^2 - \frac{1}{2L} \left( {\Phi}_{n} - {\Phi}_{n-1} \right)^2.
\end{equation}
Both capacitive and inductive energy terms are non-diagonal in position space. In compact matrix form, the Lagrangian can be written as
\begin{equation}
    \mathcal{L} = \frac{1}{2} \dot{\bm{\Phi}}^\dag\bm{V_C} \dot{\bm{\Phi}} - \frac{1}{2} \bm{\Phi}^\dag \bm{V_L\Phi},
\end{equation}
with
\begin{equation}
\begin{gathered}
    \bm{V_C}=(C_0+{2C})\mathbb{I}_N  - {C}\bm{T} - {C}\bm{T^\dag},\qquad
    \bm{V_L}=\frac{1}{L}(2\mathbb{I}_N - \bm{T} - \bm{T^\dag}).
\end{gathered}
\end{equation}
The discrete Fourier transform ${\bm U}$ simultaneously diagonalizes both matrices $\bm{V_C}$ and $\bm{V_L}$. Following the same procedure as in the previous section, the Lagrangian of the RHTL in $k$-space becomes
\begin{equation}
\begin{gathered}
    \mathcal{L}=\frac 12\sum_k\left(C_k\dot{\tilde\Phi}_k^\dag\dot{\tilde\Phi}_k-\frac 1{L_k}\tilde\Phi_k^\dag\tilde\Phi_k\right)
=\frac 12\sum_k\left(C_k\dot{\tilde\Phi}_k\dot{\tilde\Phi}_{-k}-\frac 1{L_k}\tilde\Phi_k\tilde\Phi_{-k}\right),\\
C_k  = C_0+{C}[2-2\cos(kd)],\quad L_k = \frac{L}{2-2\cos(kd)}.
\end{gathered}
\end{equation}
As before, the Hamiltonian is obtained via the Legendre transformation,
\begin{equation}
    H = \sum_k \tilde Q_k\dot{\tilde\Phi}_k-\mathcal L 
    =\sum_k \left(\frac{1}{2C_k}  \tilde Q_k \tilde Q_{-k} + \frac 1{2L_k}\tilde\Phi_k\tilde\Phi_{-k}\right).
\end{equation}
Canonical quantization promotes the dynamical variables to operators and yields a diagonal form,
\begin{equation}
    \hat H = \hbar\sum_k \omega_k \hat a_k^\dag\hat a_k.
\end{equation}
The corresponding ladder operators retain the structure given in Eq.~\eqref{eq:ladder_ope}, with a different mode-dependent impedance $Z_k = \sqrt{{L_k}/{C_k}}$. From matrices $C_{k}, L_{K}$, one directly obtains the dispersion relation,
\begin{equation}
    \omega_k = \sqrt{\frac{1}{C_k L_k}}
    =\omega_c\sqrt{\frac{1-\cos(k)}{\frac{\omega_c^2}{2\omega_0^2}+1-\cos(k)}}.
\end{equation}
where $\omega_0 = 1/\sqrt{LC_0}$ and $\omega_c = 1/\sqrt{LC}$. Using the same procedure, we obtain the coupling strength for the RHTL as,
\begin{equation}
    g_k = \frac{\hbar C_c}{2}\sqrt{\frac{\Delta}{N C_q}}\sqrt{\omega_k^3L_k}.
\end{equation}

Importantly, for this right-handed circuit the coupling strength $g_k$ cannot be expressed solely as a function of $\omega_k$. This is a direct consequence of the presence of the capacitance $C_0$, which introduces an additional $k$-dependence through $C_k$ and $L_k$. Nevertheless, in the idealized limit $C = 0$ (equivalently $\omega_c \rightarrow \infty$), one has $C_k = C_0,\; \omega_k =1/\sqrt{C_0L_k}$. In this case, the coupling strength simplifies to
\begin{equation}
    g_k = g_{0} \left(\frac{\omega_k}{\omega_{0}}\right)^{1/2}, \qquad g_0 = \frac{C_{c}}{2\sqrt{N}} \sqrt{\frac{\Delta\omega_{0}}{C_{q} C_{0}}},
\end{equation}
which recovers the familiar $\sqrt{\omega_k}$ scaling \cite{USC_Open_Lines2013}.
%
\section{Self-energy: Decay Rate and Lamb Shift}
\begin{figure}[t!]
    \centering
    \includegraphics[width=0.5\linewidth]{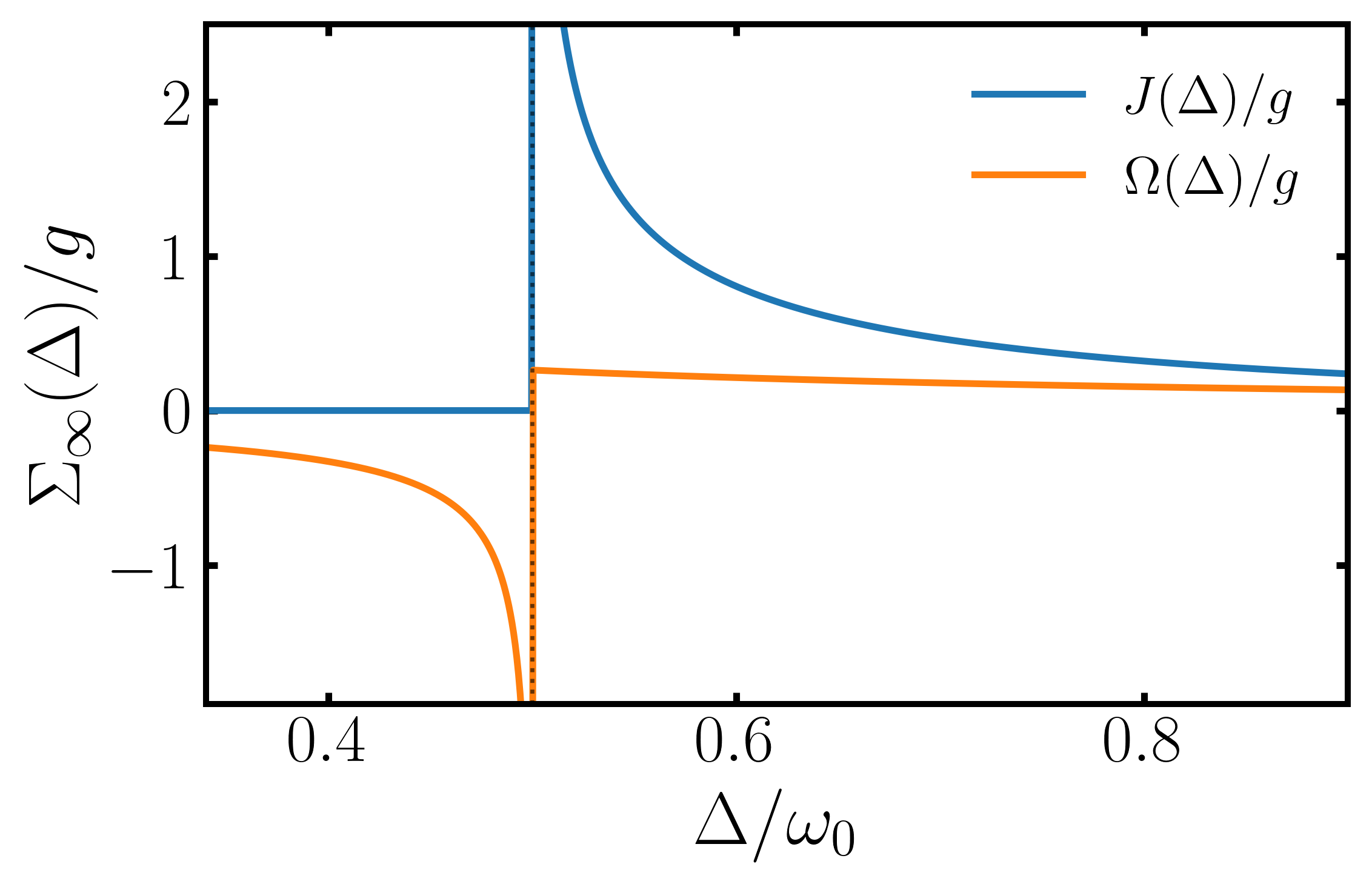}
    \caption{ \textbf{Self-energy function.} Lamb shift \(\Omega(\Delta)\) and spectral function \(J(\Delta)\) extracted from real and imaginary parts of the self-energy in the infinite-UV-cutoff limit. The vertical dotted line at \(\Delta=\omega_0/2\) denotes the LHTL lower band edge, where both quantities diverge.}
    \label{fig: Self-energy for infinite cutoff}
\end{figure}

%
In this section, we report the closed form of self-energy for the qubit-LHTL system obtained in the infinite UV-cutoff limit ($\omega_c \to \infty$) in the continuum limit ($N \to \infty$). To this end, we follow the standard procedure of first calculating the bound-state energy and consider the general single-excitation eigenstate of the full Hamiltonian,
\begin{equation}
    |\psi_{\mathrm{bound}}\rangle
    =
    C_e\,\sigma^+ |g,\mathbf{0}\rangle
    +
    \sum_k C_k\,a_k^\dagger |g,\mathbf{0}\rangle,
    \label{eq:boundwf}
\end{equation}
where $C_e$ and $C_k$ denote the emitter and photonic amplitudes, respectively. Requiring the bound state to satisfy the time-independent Schr\"odinger equation $H|\psi_{\mathrm{bound}}\rangle
    =
    E_{\mathrm{bound}}|\psi_{\mathrm{bound}}\rangle$, 
and projecting onto the states $|e,\mathbf{0}\rangle$ and $a_k^\dagger |g,\mathbf{0}\rangle$ yields the coupled equations
\begin{equation}
    C_e = \frac{1}{E_{\mathrm{bound}}-\Delta}\sum_k g_k\,C_k,
    \qquad
    C_k = \frac{C_e\,g_k^*}{E_{\mathrm{bound}}-\omega_k}.
\end{equation}
Real solutions to $E_{\rm bound} \in \mathbb{R}\setminus [\omega_{\rm IR}, \omega_{\rm UV}]$ determine the existence of qubit-photon bound-states, obtained from the self-consistent equation,
\begin{equation}\label{eq: Bound-state equation}
    E_\text{bound} = \Delta + \Sigma(E_\text{bound}), \hspace{5mm}\Sigma(E) = \sum_k \frac{|g_k|^2}{E - \omega_k},
\end{equation}
where $\Sigma(E)$ is the self-energy function of the system,
\begin{equation}
    \Sigma(E) = \sum_k \frac{|g_k|^2}{E- \omega_k}, \qquad g_k \simeq \frac{g}{\sqrt{N}}
\end{equation}
A closed form of $\Sigma(E)$ can be calculated in the continuum limit ($N \to \infty$) all $k$-sums are expressed in terms of the following integral, 
\begin{equation}
    \lim_{N \to \infty} \frac{1}{N}\sum_k \to  \int\,\frac{dk}{2\pi}.
\end{equation}
For infinite cutoff $\omega_c \to \infty$ case one obtains closed function forms, 
\begin{equation}
   \begin{split}
        \Sigma_\infty(E) &= \frac{g^2}{2\pi}\int_{-\pi}^\pi\,\frac{dk}{E - \omega_0/2|\sin(k/2)|} \\
        &= \begin{cases}
           \dfrac{g^2}{E}\left[1+\dfrac{4\omega_0}{\pi\sqrt{\omega^2_0-4E^2}}\tan^{-1}\left(\sqrt{\frac{\omega_0+2E}{\omega_0-2E}}\right)\right]; &E<\omega_0/2\\*\\[0pt]
            \dfrac{g^2}{E}\left[1+\dfrac{2\omega_0}{\pi\sqrt{4 E^2-\omega_0^2}}\left\{\ln\left({\frac{2E+\sqrt{4E^2-\omega_0^2}}{\omega_0}}\right)-i\pi\right\}\right], &E>\omega_0/2
        \end{cases}
   \end{split}
\label{eq:selfenergy}
\end{equation}
In general the self-energy of a system is complex-valued and its real and imaginary parts correspond to frequency shift and decay of the qubit, see Fig.~\ref{fig: Self-energy for infinite cutoff}. More precisely, the retarded self-energy is written as, 
\begin{equation}
    \Sigma_\infty^R(E + i 0^+) = \Omega(E) - \frac{i J(E)}{2}
\end{equation}
where $\Omega(E) = \text{Re}(\Sigma_\infty^R(E + i 0^+))$ is the Lamb's shift, and $J(E) = -2\text{Im} (\Sigma_\infty^R(E + i 0^+))$ is the spectral function corresponding to the effective decay rate of qubit into the waveguide.
%
\section{Algebraic Bound-state}
%
In this section, we derive the bound-state profile for both left-handed and right-handed transmission lines. A bound state arises when the dressed eigenenergy of the coupled emitter--waveguide system lies outside the photonic continuum, such that the self-energy is purely real and radiative decay into propagating waveguide modes is forbidden. In this regime, an excitation initially localized on the emitter hybridizes with evanescent photonic modes of the waveguide to form a stationary dressed state shared by the emitter and the field.
\par
Rewriting the bound-state wavefunction of Eq.~(\ref{eq:boundwf}) in the form
\begin{equation}
    |\psi_{\mathrm{bound}}\rangle
    =
    C_e
    \left(
    \sigma^+
    +
    \sum_k
    \frac{g_k^*}{E_{\mathrm{bound}}-\omega_k}\,a_k^\dagger
    \right)
    |g,\mathbf{0}\rangle,
\end{equation}
and expressing the momentum-space creation operators in terms of real-space photonic operators.
\begin{equation}
    \begin{split}
        |{\psi_\text{bound}}\rangle &= C_e\, \left( \sigma^+ + \sum_k \frac{g_k^*}{E_\text{bound} - \omega_k} \frac{1}{\sqrt{N}} \sum_n e^{-ikn} a_n^\dagger \right) \,|{g,\textbf{0}}\rangle\\
        &= C_e\, \left( \sigma^+ + \sum_n \underbrace{\frac{g}{N} \sum_k \frac{e^{-ik(n-m)}}{E_\text{bound}-\omega_k}}_{\phi_{n-m}} a_n^\dagger \right) \,|{g,\textbf{0}}\rangle
    \end{split}
\end{equation}
we can obtain the spatial structure of the photonic component (bound-state profile). To characterize the photonic amplitude as a function of distance from the emitter, we set $m=0$ to evaluate the photon occupation at a site $n$ relative to the emitter as,
\begin{equation}\label{eq: bound-state profile integral}
\begin{split} 
    \langle a_n^\dagger a_n \rangle_{\rm bound} = |C_{e}|^{2} |\phi_{n}|^{2} 
\end{split}
\end{equation}
where
\begin{equation}\label{SMeq: phi_n_general}
    \phi_n
    =
    \frac{g}{2\pi}\int_{-\pi}^{\pi} dk\,
    \frac{e^{-ikn}}{E_{\mathrm{bound}}-\omega(k)}
\end{equation}
is evaluated in the continuum-$k$ limit $(N\to \infty)$ such that the unit-cell size, and consequently $k_{\rm max}$, remain finite. The normalization $|C_e|^2$ is set by demanding a normalized bound-state wavefunction
\begin{equation}
    \langle \psi_{\mathrm{bound}} | \psi_{\mathrm{bound}} \rangle
    =
    |C_e|^2
    \left(
    1
    +
    \sum_k
    \frac{|g_k|^2}{(E_{\mathrm{bound}}-\omega_k)^2}
    \right) = 1.
\end{equation}
From the self-energy function $\Sigma(E)=\sum_k {|g_k|^2}/{(E-\omega_k)}$, $C_e$ can be written in the compact form
\begin{equation}
    |C_e|^2
    =
    \left[
    1-\frac{d\Sigma(E)}{dE}\bigg|_{E=E_{\mathrm{bound}}}
    \right]^{-1}.
\end{equation}
\par
The asymptotic form of the bound-state profile is determined by the long-wavelength (low-$k$) sector of the dispersion $\omega(k)$, since for $n\gg 1$ the oscillatory factor $e^{-ikn}$ suppresses contributions from high-momentum modes.
\begin{figure}[!t]
    \centering
    \includegraphics[width=0.45\linewidth]{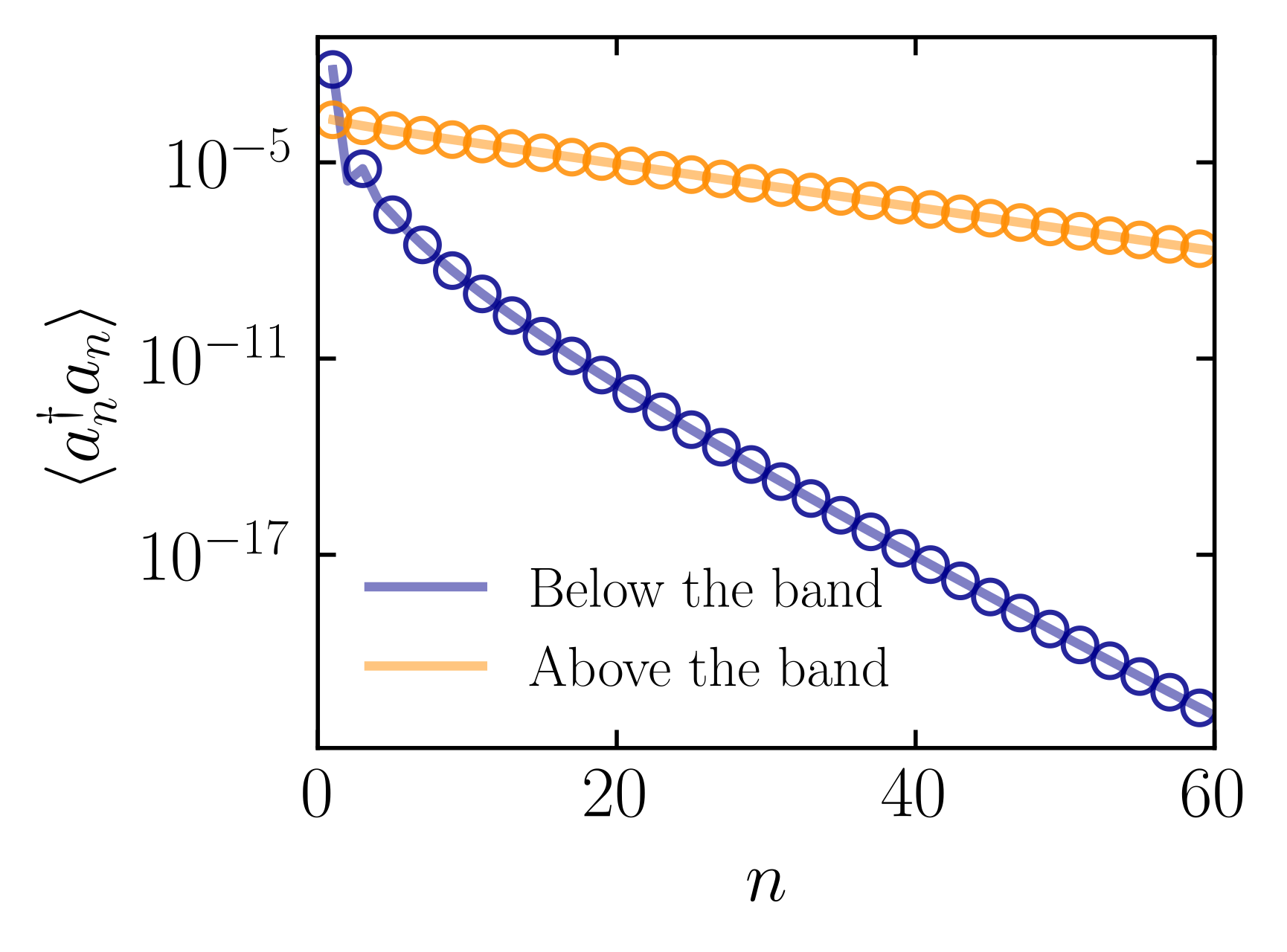}
    \caption{\textbf{Numerical validation of bound states.} Bound-state profiles for an emitter tuned below and above the LHTL photonic band are shown. Solid lines represent the analytical result from Eq.~\eqref{eq: bound-state profile integral}, while circular markers denote exact diagonalization of the single-excitation Hamiltonian in Eq.~\eqref{SMeq: left-handed waveguide-QED Hamiltonian} following the procedure described in Sec.~\ref{sec: Numerical Diagonalization}. Parameters: $g/\omega_0 = 0.08$, $\Delta_{\mathrm{below}}/\omega_0 = 0.3$, $\Delta_{\mathrm{above}}/\omega_0 = 4.2$, $\omega_c/\omega_0 = 4$, and $N = 4001$.}
    \label{fig: Numerical Validation}
\end{figure}
For the LHTL, the low-$k$ dispersion is strongly enhanced and may be approximated as
\begin{equation}
    \omega(k\approx 0)
    \simeq
    \frac{\omega_c}{\sqrt{1+(\omega_c k/\omega_0)^2}}.
\end{equation}
Since the bound-state energy lies well \emph{below} this large low-$k$ scale, the denominator in Eq.~\eqref{SMeq: phi_n_general} is dominated by $\omega(k)$, and we may approximate
\begin{equation}
    \frac{1}{E_{\mathrm{bound}}-\omega(k)}
    \simeq
    -\frac{1}{\omega(k)}.
\end{equation}
The long-distance photonic amplitude then becomes
\begin{equation}
    \phi_n
    \simeq
    -\frac{g}{2\pi \omega_c}
    \int_{-\infty}^{\infty} dk\,
    e^{-ikn}\sqrt{1+(\omega_c k/\omega_0)^2},
\end{equation}
where the integration limits have been extended to $\mathbb{R}$, since the dominant contribution comes from the long-wavelength region. This Fourier transform can be evaluated in closed form, yielding
\begin{equation}
    \phi_n
    \simeq
    -\frac{g}{\pi \omega_c}\,
    \frac{K_1(n\, \omega_0/\omega_c)}{n},
\end{equation}
and, therefore,
\begin{equation}
    \langle a_n^\dagger a_n\rangle
    \simeq
    \frac{g^2 |C_e|^2}{\pi^2 \omega_0^2 n_\star^2}
    \left[\frac{K_1(n/n_\star)}{n}\right]^2, \qquad n_\star \equiv \frac{\omega_\text{UV}}{2\omega_\text{IR}} \simeq \frac{\omega_c}{\omega_0}.
\label{eq:boundprofile}
\end{equation}
Here $K_1(z)$ is the modified Bessel function of the second kind. This analytical result is consistent with the numerical results obtained via an exact diagonalization of single-excitation Hamiltonian. Interestingly, when the bound-state energy is sufficiently far away from the band-edge, this results remains surprisingly accurate even for relatively small $n$. Equation~(\ref{eq:boundprofile}) makes it clear that the LHTL bound state profile exhibits a crossover over length $n_\star$: the photonic cloud is neither purely exponential nor set by a single lattice-scale decay length, but instead inherits its spatial structure directly from the anomalous low-$k$ dispersion of the waveguide.
\par
Interestingly, for finite $\omega_c$, an emitter tuned above the LHTL band forms a bound state that is much more extended than the one formed below the band, even when both are kept at equal energy detuning from their respective band edges (see Fig.~\ref{fig: Numerical Validation}). This asymmetry originates from the fact that the two edges are of different momentum sectors: the upper bound state is predominantly built from $k \approx 0$ long-wavelength modes, whereas the lower bound state is dominated by $k \approx \pi$ short-wavelength modes. Because the $k \approx 0$ modes vary slowly in space, they produce a much broader real-space envelope, while the $k \approx \pi$ modes lead to a more tightly localized bound state.\newline
\par 
A similar analysis applies to the RHTL. In this case the dispersion is gapless,
\begin{equation}
  \omega_k^{\mathrm{RH}}
  =
  \omega_c
  \sqrt{
  \frac{1-\cos k}
  {\frac{\omega_c^2}{2\omega_0^2}+1-\cos k}
  },
\end{equation}
and the bound state is formed for an emitter frequency above the band. The long-distance behavior is again governed by the low-$k$ sector, where $\omega_k^{\mathrm{RH}}\simeq \omega_0 |k|$. Since $E_{\mathrm{bound}}\gg \omega(k)$ in this regime, we expand
\begin{equation}
    \frac{1}{E_{\mathrm{bound}}-\omega(k)}
    \simeq
    \frac{1}{E_{\mathrm{bound}}}
    +
    \frac{\omega(k)}{E_{\mathrm{bound}}^2}.
\end{equation}
For $n\neq 0$, the constant term contributes only to the on-site component and does not affect the spatial tail. The leading off-site behavior is therefore set by the term proportional to $\omega(k)\propto |k|$, whose Fourier transform gives an amplitude decaying as $1/n^2$. As a result, the photon density falls off as
\begin{equation}
    \langle a_n^\dagger a_n\rangle
    \simeq
    \frac{\omega_0^2 g^2 |C_e|^2}{\pi^2 E_{\mathrm{bound}}^4}
    \frac{1}{n^4}.
\end{equation}
Thus, while both transmission lines support bound states outside the continuum, the spatial structure of the photonic cloud is qualitatively different: the LHTL profile is governed by the crossover scale $n_\star$, whereas the RHTL exhibits the familiar algebraic $1/n^4$ decay set by its linear low-energy dispersion.

Inclusion of generic momentum-dependent $g(k)$ modifies the bound-state profiles for LHTL and RHTL as,
\begin{equation}\label{eq: Exact gk LH Bound-state}
    \begin{split}
        \langle a_n^\dagger a_n \rangle^{\rm LH} &= \frac{\omega_0 g_0^2 |C_e|^2}{4 \pi^{5/2} n_\star^{3/2} \varpi} \left( \frac{K_{1/4}(n/n_\star)}{n^{1/4}} \right)^2,
       \quad \langle a_n^\dagger a_n \rangle^{\rm RH} = \left(\frac{\omega_0 g_0^2 |C_e|^2}{{16 \pi^2 E_\text{bound}^2}}\right) \frac{1}{n^{3}}
    \end{split}
\end{equation}
where $\varpi \approx 2.622$ is the Lemniscate constant that commonly appears while evaluating Elliptic integrals. 
%
%
\section{Bound-state core: waveguide-QED near the band-edge}
\label{subsec: Near band-edge exponential drop}
%
\begin{figure}[!th]
    \centering
    \includegraphics[width=0.4\linewidth]{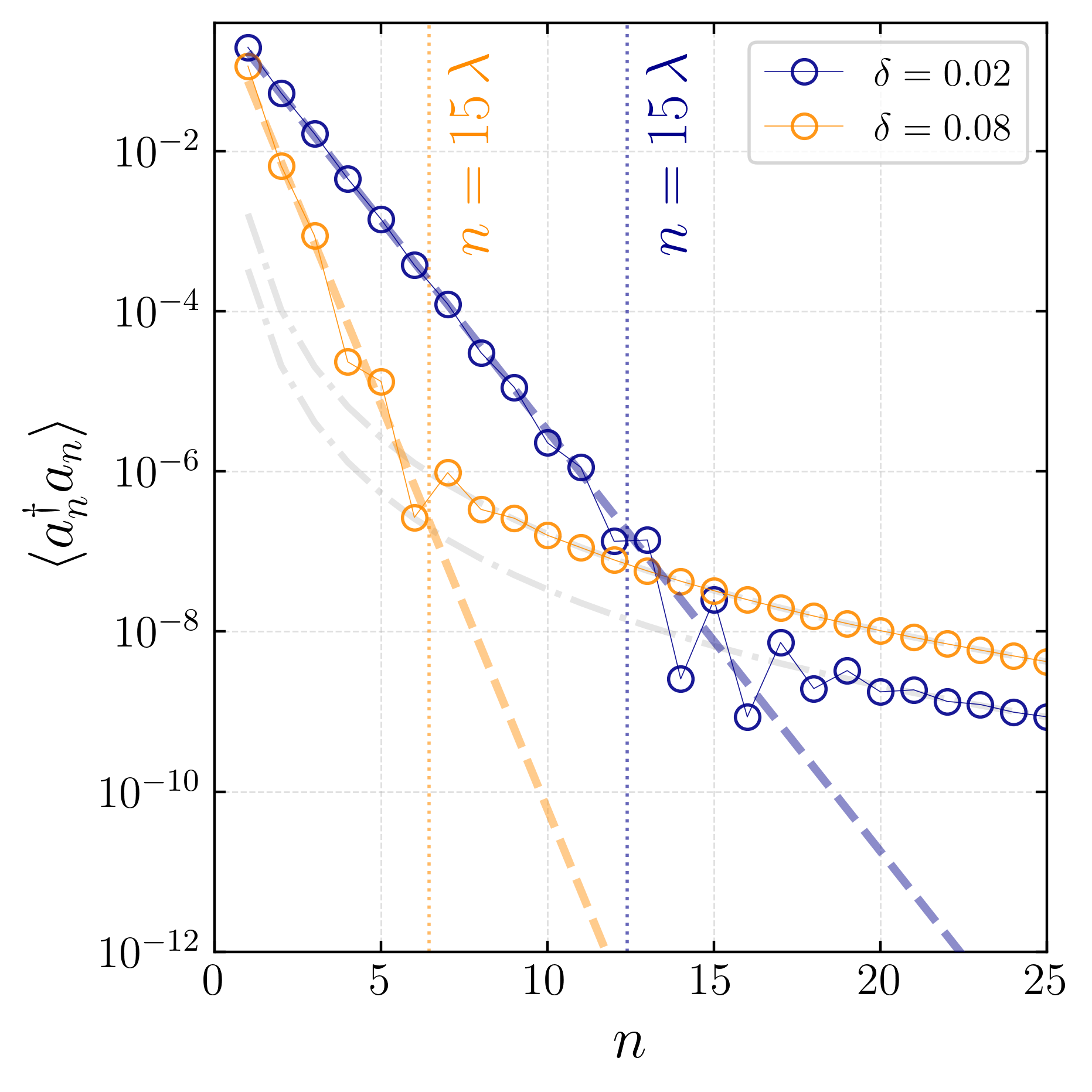}
    \caption{\textbf{LHTL bound-state core exponential window.} Colored dashed curves show the short-distance exponential core of the LHTL bound-state profile for two detunings, \(\delta=\omega_{\mathrm{IR}}-E_{\mathrm{bound}}\). The vertical dotted lines delineate the approximate extent of the exponential window. At larger distances, the profile crosses over to the asymptotic algebraic tail, shown by the gray dash-dotted curve, with \(\langle a_n^\dagger a_n\rangle \sim 1/n^4\). Parameters: \(\omega_0=1\), \(g/\omega_0=0.2\), and \(\omega_c\to\infty\).}
    \label{fig: LHTL Bound-state core}
\end{figure}
For the LHTL, when the qubit is tuned just below the lower band edge, it is useful to define the detuning from the band edge as $\delta \equiv \omega_{\rm IR} - E_{\rm bound}$. In addition to the algebraic-to-exponential crossover governed by $n_\star$, the bound-state profile also exhibits a short-distance exponential core. This near-field structure is controlled by the quadratic form of the dispersion near the band minimum at $k \simeq \pi$. For simplicity, we work in the infinite-cutoff limit. Writing $q=\pi-k$, the LHTL dispersion becomes
\begin{equation}
    \omega(k)\simeq \frac{\omega_0}{2}\left(1+\frac{q^2}{8}\right),
\end{equation}
so that the bound-state kernel may be approximated as
\begin{equation}
    \frac{1}{E_{\rm bound}-\omega(k)}
\simeq
-\frac{1}{\delta + (\omega_0/16)q^2}.
\end{equation}
In this regime, the approximation $E_{\rm bound}-\omega(k)\approx -\omega(k)$, used in the previous section to capture the long-distance tail, is no longer valid. Instead, the Fourier integral is dominated by momenta near the band edge and yields, for separations $|n|\gtrsim 1$,
\begin{equation}
\langle a_n^\dagger a_n \rangle_{\rm LH}
\approx
\frac{4 g^2 |C_e|^2}{\omega_0 \delta}\,
e^{-|n| / \lambda_{\rm LH}},
\qquad
\lambda_{\rm LH} \equiv \frac{1}{\sqrt{2}}\sqrt{\frac{\omega_0/4}{\delta}}
= \frac12 \sqrt{\frac{\omega_0}{2\delta}}.
\end{equation}
This identifies a core localization length $\lambda_{\rm LH}$ for the intensity profile. The exponential core is confined to the near field, $n \lesssim \mathrm{few}\times \lambda_{\rm LH}$, before crossing over to the broader profile discussed in previous section; see Fig.~\ref{fig: LHTL Bound-state core}. In the natural regime $\delta \gg \omega_0/64$, one has $\lambda_{\rm LH}\ll 1$, so this near-field exponential structure is essentially restricted to the first unit cell.

A similar analysis may be carried out for the RHTL. In that case one finds
\begin{equation}
\langle a_n^\dagger a_n \rangle_{\rm RH}
\approx
\frac{g^2 |C_e|^2}{\omega_0 \delta}\,
e^{-|n| / \lambda_{\rm RH}},
\qquad
\lambda_{\rm RH} \equiv \frac14 \sqrt{\frac{\omega_0}{\delta}}.
\end{equation}
Thus, in both transmission lines, the short-distance core is exponentially localized with a characteristic length that scales as $\lambda \propto \delta^{-1/2}$. Finite ultraviolet cutoffs modify these expressions only weakly, provided the system remains in the good-waveguide regime, $\omega_c > \omega_0$.
%
\section{Reproducing the dispersion by extended hopping}
%
%
\begin{figure}[b!]
    \centering
    \includegraphics[width=0.5\linewidth]{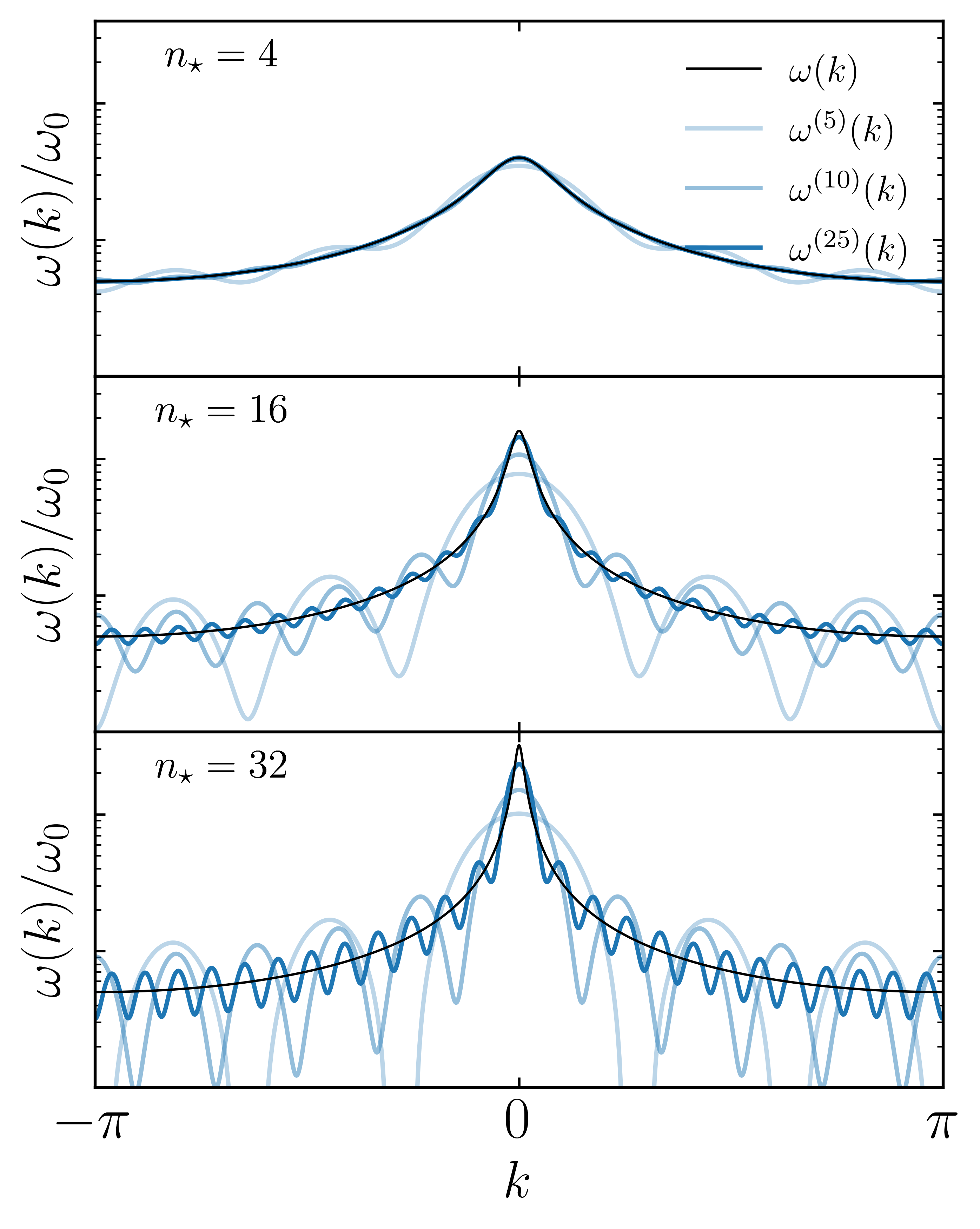}
    \caption{\textbf{Dispersion relation.} Solid blue lines represent the dispersion relations obtained using Eq.~\eqref{SMeq: truncated dispersion} for $D = 5,10,25$. Black solid lines denote the exact dispersion.}
    \label{fig: Truncated Hamiltonian dispersion}
\end{figure}
Writing LHTL Hamiltonian in the Wannier basis, where the cosine Fourier coefficients of the dispersion $\omega_k$ act as photon hopping amplitudes between sites separated by $n$ unit cells, \cite{Moghaddam2021}, the effective real-space photonic network associated with the LHTL can be expressed as,
\begin{equation}
\label{SMeq:LHTL-hopping}
\begin{split}
    H_{{\rm LH}}
    &= \sum_{m\in\mathbb{Z}}\sum_{n\in\mathbb{Z}} \xi_n\, a_{m+n}^\dagger a_m,\\
    \xi_n
    &= \frac{1}{\pi}\int_{0}^{\pi}\omega(k)\cos(nk)\,dk,
\end{split}
\end{equation}
where $\xi_n$ is the hopping amplitude between sites separated by $n$ unit cells. This representation makes explicit that the LHTL is not described by a short-range tight-binding model, but by an extended hopping network whose range is set by the structure of the waveguide dispersion.

To quantify the role of long-range hopping, we introduce a truncated Hamiltonian $H_{\rm LH}^{(D)}$, in which photonic hopping is retained only up to a distance $D$,
\begin{equation}
    H_{\rm LH}^{(D)}
    =
    \sum_{m\in\mathbb{Z}}\sum_{n=-D}^{D}\xi_n\,a_{m+n}^\dagger a_m.
\end{equation}
Its corresponding dispersion is
\begin{equation}\label{SMeq: truncated dispersion}
    \omega^{(D)}(k)
    =
    \xi_0 + 2\sum_{n=1}^{D}\xi_n\cos(kn).
\end{equation}
Eq.~\eqref{SMeq: truncated dispersion} shows that truncating the real-space hopping is equivalent to approximating the exact dispersion by a finite cosine series. As shown in Fig.~\ref{fig: Truncated Hamiltonian dispersion}, this approximation fails to capture the characteristic low-$k$ structure of the LHTL unless the hopping cutoff extends well beyond the crossover scale, $D \gg n_\star$. In other words, the anomalous long-wavelength dispersion of the LHTL is encoded in a genuinely long-range hopping network in real space, which cannot be reproduced by a finite-distance truncation.
%
\section{Exact Qubit Dynamics}
%
In this section, we derive the exact qubit dynamics $C_e(t)$ for an initially excited emitter coupled to the waveguide, emphasizing the role of the structured LHTL bath in generating non-Markovian behavior. We first reduce the coupled emitter--photon problem to an integro-differential equation for the qubit amplitude, whose kernel is determined by the waveguide spectral density. We then solve this equation by inverse Laplace transformation and analyze the singularity structure of the resulting integrand in the complex plane, following Ref.~\cite{Berman2010}. This approach naturally separates the dynamics into a branch-cut contribution associated with the continuum of propagating modes and a pole contribution associated with the bound state, thereby capturing within a single framework both the near-exponential decay at intermediate times and the algebraic long-time relaxation set by the band edge.

We begin with the time-dependent ansatz
\begin{equation}\label{SMeq: Time-dependent state}
    |\psi(t)\rangle = C_e(t)|e,0\rangle + \sum_k C_k(t)\,a_k^\dagger|g,0\rangle,
\end{equation}
with initial conditions $C_e(0)=1$ and $C_k(0)=0$, where $C_e(t)$ denotes the emitter amplitude and $C_k(t)$ the photonic amplitude. Substituting this ansatz into the time-dependent Schr\"odinger equation yields the coupled system of equations
\begin{subequations}
    \begin{align}
        \Delta\, C_e(t) + \sum_k g_k\,C_k(t) &= i \dot C_e(t),\\
        g_k^*\,C_e(t) + \omega_k\,C_k(t) &= i \dot C_k(t).
    \end{align}
\end{subequations}
It is convenient to calculate these amplitudes in the interaction frame by redefining the amplitudes as,
\[
\tilde C_e(t) \equiv e^{i\Delta t} C_e(t),
\qquad
\tilde C_k(t) \equiv e^{i\omega_k t} C_k(t),
\]
in terms of which the equations become
\begin{subequations}
    \begin{align}
        i \dot{\tilde C}_e(t) &= \sum_k g_k\,e^{-i(\omega_k-\Delta)t}\,\tilde C_k(t), \label{subeq: C_e diff equation}\\
        i \dot{\tilde C}_k(t) &= g_k^*\,e^{i(\omega_k-\Delta)t}\,\tilde C_e(t). \label{subeq: C_k diff equation}
    \end{align}
\end{subequations}
Eliminating the photonic amplitudes $\tilde C_k(t)$, we obtain the integro-differential equation for emitter amplitude
\begin{equation}\label{SMeq: tilde C_e diff equation SM}
    \dot{\tilde C}_e(t)
    =
    -\int_0^t K(\tau)\,\tilde C_e(t-\tau)\,d\tau,
\end{equation}
where $K(\tau) = (2\pi)^{-1}
\int_0^\infty {d\omega}\,J(\omega)\,e^{-i(\omega-\Delta)\tau}$ is the memory kernel set by the spectral density $J(\omega)$. We recast Eq.~\eqref{SMeq: tilde C_e diff equation SM} as an algebraic equation by taking its Laplace transform, 
\begin{equation}
    \tilde C_e(s) = \frac{1}{s+K(s)},
\end{equation}
where the transform is defined as,
\begin{equation}
    f(s) \equiv \int_0^\infty e^{-st}f(t)\,dt.
\end{equation}
The transformed memory kernel can be written as
\begin{equation*}
    K(s)
    =
    \int_0^\infty \frac{d\omega}{2\pi}\,
    \frac{J(\omega)}{s+i(\omega-\Delta)}
    =
    i\,\Sigma(is+\Delta), \qquad \Sigma(z)=\int_0^\infty d\omega\,\frac{J(\omega)}{z-\omega},
\end{equation*}
where $\Sigma(z)$ is the self-energy. The qubit amplitude then follows from the inverse Laplace transform of
\[
\tilde C_e(s)=\bigl[s+i\Sigma(is+\Delta)\bigr]^{-1},
\]
which leads to
\begin{equation}
    C_e(t)
    =
    \frac{1}{2\pi i}
    \int_{\sigma-i\infty}^{\sigma+i\infty}
    \frac{e^{\tilde s t}\,d\tilde s}{\tilde s+i\Delta+G(\tilde s)},
\end{equation}
with $\tilde s=s-i\Delta$ and $G(\tilde s)=i\Sigma(i\tilde s)$. The integrand has a branch cut along $\tilde s \in [-i\omega_0/2,-i\infty)$, so the Bromwich contour is deformed to avoid crossing it, as shown in Fig.~4A of the main text. The deformed contour also encloses a bound-state pole at $\tilde s=i y_p$, where $y_p$ satisfies
\[
y-\Delta-\Sigma(-y)=0.
\]
As a result, the dynamics separates into a continuum contribution arising from the discontinuity across the branch cut and a bound-state contribution from the pole residue:
\begin{equation}
    C_e(t)
    =
    C_e^{(s)}(t)+C_e^{(b)}(t)
    =
    \frac{8g^2}{\pi\omega_0^2}\int_{-\infty}^{-1} dx\,M(x)\,e^{i\omega_0 x t/2}
    +
    r\,e^{i y_p t}.
\end{equation}
Here the spectral weight function is
\begin{equation}
    M(x)=
    \frac{-x\sqrt{x^2-1}}
    {x^2(x^2-1)\left[x+\frac{2\Delta}{\omega_0}
    -\frac{4g^2}{\omega_0^2 x}
    \left(1+\frac{2\ln\left(\sqrt{x^2-1}-x\right)}{\pi\sqrt{x^2-1}}\right)\right]^2
    +\left(2\sqrt{2}g/\omega_0\right)^4},
\end{equation}
where $x\equiv 2y/\omega_0$ is a dimensionless energy variable. The residue of the pole is $r = \left[1+G'(\tilde s)\right]^{-1}_{\tilde s=i y_p},$ with $G'(z)\equiv dG/dz$.
%
\subsubsection{Exponential decay}
%
For early times, the qubit decays into the spectrally closest modes available in the LHTL spectrum around which the weight function $M(x)$ is approximately Lorentzian. Thus, the major peak of $M(x)$ around the qubit frequency dictates its early-time dynamics. With the pole location in the complex-plane (which is same as peak location in real-space with a width corresponding to the imaginary part of pole location) denoted by $x_0 \in \mathbb{C}$, we approximate $M(x)$ around $x = x_0$ as,
\begin{equation}
    \begin{split}
        M(x) &= \frac{N(x)}{D(x)} = \frac{N(x_0)}{D(x_0)+D'(x_0)(x-x_0) + D''(x_0)(x-x_0)^2/2+\dots} \\
        &\approx \frac{N(x_0)/D'(x_0)}{(x-x_0)} + \frac{N(x_0^*)/D'(x_0^*)}{(x-x_0^*)} \\
    &= \frac{a_0}{x-x_0} + \frac{a_0^*}{x-x_0^*},\hspace{30mm} a_0 \equiv \frac{N(x_0)}{D'(x_0)}
    \end{split}
\end{equation}
where the denominator $D(x_0) = 0$, and we assume $D'(x_0) \neq 0$. There is a pair of complex conjugate poles at $x_{0}$ whose location can be calculated numerically by solving the following equation,
\begin{equation}
    x_0^2(x_0^2-1)\left[x_0+\frac{2\Delta}{\omega_0} - \frac{4g^2}{\omega_0^2x_0}\left( 1+\frac{2\,\ln\left(\sqrt{x_0^2-1}-x_0\right)}{\pi \sqrt{x_0^2-1}}\right)\right]^2+\left(2\sqrt{2}g/\omega_0\right)^4 = 0.
\end{equation}
We evaluate the emitter amplitude $C_e(t)$ as, 
\begin{equation}
    \begin{split}
         C_e^{(s)}(t) &\approx -\frac{8g^2a_0}{\pi \omega_0^2}\int_{-\infty}^{+\infty}\,dx\,\frac{e^{i\omega_0 x t/2}}{x-x_0}\\
         &=  -\frac{16ig^2a_0}{\omega_0^2}\,e^{-\gamma t/2}\,e^{i \delta t}
    \end{split}
\end{equation}
where, $\gamma = \omega_0 \,\Im(x_0)$, and $\delta = \Re(x_0)\,\omega_0/2$. Note that we have extended the limits of the integral from $-\infty$ to $+\infty$ and close the contour above the real axis enclosing only the pole in the upper-half of the complex plane to ensure convergence (i.e. $\gamma > 0$). The emitter population then reads, 
\begin{equation}\label{SMeq: Intermediate time qubit population, exponential decay}
    |C_e^{(s)}(t)|^2 \approx \left( \frac{16 g^2 |a_0|}{\omega_0^2} \right)^2\,e^{-\gamma t}
\end{equation}
%
\subsubsection{Polynomial $t^{-3}$ tail}
%
At long times, $t \gg 1/\gamma$, the continuum contribution is governed by the non-analytic structure of the $M(x)$ near the lower band edge, $x=-1$. The asymptotic dynamics therefore follow from the behavior of $M(x)$ in the vicinity of this branch point, where it takes the form
\begin{equation}
    M_1(x)\approx \frac{1}{32\sqrt{2}}\left(\frac{\omega_0}{g}\right)^4 \sqrt{-1-x},
    \qquad x<-1.
\end{equation}
To evaluate the branch-cut integral in closed form, we regularize this edge approximation by introducing a soft exponential cutoff,
\begin{equation}
    \tilde M(x)\equiv e^{x+1}M_1(x)
    =
    \frac{1}{32\sqrt{2}}\left(\frac{\omega_0}{g}\right)^4
    e^{x+1}\sqrt{-1-x},
\end{equation}
which preserves the square-root singularity at $x=-1$ and therefore does not modify the long-time asymptotic behavior. Substituting $\tilde M(x)$ into the continuum contribution yields
\begin{equation}
    \begin{split}
        C_e^{(s)}(t)
        &\approx
        -\frac{8g^2}{\pi \omega_0^2}\,
        \frac{1}{32\sqrt{2}}
        \left(\frac{\omega_0}{g}\right)^4
        \int_{-\infty}^{-1} dx\,
        e^{i\omega_0 x t/2}\,
        e^{x+1}\sqrt{-1-x} \\
        &=
        -\frac{\omega_0^2}{8\sqrt{2\pi}\,g^2}\,
        \frac{e^{-i\omega_0 t/2}}{(1+i\omega_0 t/2)^{3/2}}.
    \end{split}
\end{equation}
Accordingly, the excited-state population exhibits the algebraic long-time decay
\begin{equation}\label{eq: long time qubit population, polynomial decay}
    |C_e^{(s)}(t)|^2
    \simeq
    \frac{1}{16\pi}
    \left(\frac{\omega_0}{g}\right)^4
    \frac{1}{(\omega_0 t)^3},
    \qquad \omega_0 t \gg 1.
\end{equation}
Thus, the $t^{-3}$ tail arises directly from the square-root band-edge singularity of the LHTL spectrum and constitutes the universal asymptotic form of the continuum contribution.
%
\section{Light-cone expression}
%
To characterize the spreading of an initially localized excitation, we extract an $\epsilon$-light-cone from the site-resolved photon signal. For each site $n$, we define the arrival time $t_\epsilon(n)$ as the earliest time at which the local response exceeds a fixed threshold $\epsilon$, using the same threshold for all sites. Introducing the dimensionless distance variable $z \equiv n/n_\star$, we equivalently work with the arrival-time profile $t_\epsilon(z) \equiv t_\epsilon(n=z\,n_\star)$.\newline
\par 
To this end, we calculate the spatial amplitude of scattered photons as the inverse Fourier transform of $C_k(t)$ (see Eq.~\ref{subeq: C_k diff equation}), as
\begin{equation}
    \begin{split}
        C_n(t) &= \frac{g}{2\pi} \int_{-\pi}^\pi\,dk\,e^{ikn}\,\left( \int_0^t d\tau e^{-i(t-\tau)\omega(k)} C_e(\tau) \right) \\
    &= g\!\int_{0}^{t} d\tau\, G^R_{n}(t-\tau)\,C_e(\tau)
    \end{split}
\end{equation}
where in the second equation, the photon amplitude has been recast into an manifestly time-delayed form using the retarded photon propagator,
\begin{equation}
\begin{split}
    G^R_n(t) &\equiv -i\,\Theta(t)\,\langle[a_n(t),a_0^\dagger(0)]\rangle\\
    &= {(2\pi)^{-1}}\!\int_{-\pi}^{\pi}\,dk\,e^{ikn-i\omega(k) t}.
\end{split}
\end{equation}
At very short-times we approximate $C_e(\tau) \approx 1$, and expand
$G^R_{n}(t)\simeq-i\xi_n t+O(t^2)$ to obtain
\begin{equation}
    C_n(t) \approx -\frac{i g}{2}\,\xi_n\, t^2,
\end{equation}
so that the local signal inherits its spatial dependence directly from the real-space hopping kernel $\xi_n$. Consequently, the arrival time scales as
\begin{equation}\label{SMeq: short-time photon dynamics}
    t_\epsilon(z) \propto \frac{1}{\sqrt{\xi(z)}},
\end{equation}
where the proportionality constant depends on the $\epsilon$ but is independent of distance. Thus, the $\epsilon$-cone at short times is governed entirely by the spatial decay of the hopping network.\newline
\par
Following the running-exponent construction described under Methods, we parametrize the $\epsilon$-cone in terms of a position-dependent light-cone exponent $\gamma(z)$,
\begin{equation}
t_\epsilon(z)=t_\epsilon(z_0)\left(\frac{z}{z_0}\right)^{-\bar{\gamma}(z;z_0)},
\qquad
\bar{\gamma}(z;z_0)=\frac{1}{\ln(z/z_0)}
\int_{z_0}^{z}\frac{\gamma(z')}{z'}\,dz'.
\label{eq:gamma-bar-def}
\end{equation}
Comparing Eq.~\eqref{SMeq: short-time photon dynamics} with the definition of the local hopping exponent, $\alpha(z) \equiv - d\ln \xi(z)/d\ln z$, immediately gives
\begin{equation}
    \gamma(z)=\frac{\alpha(z)}{2},
    \qquad \text{(short-time regime)}
\end{equation}
and consequently $\bar\gamma(z;z_0)=\bar\alpha(z;z_0)/2$. To obtain an explicit $\epsilon$-cone, we set the initial value $t_\epsilon(z_0 = 1/n_\star)$ using a numerical estimation of arrival time at which the light front amplitude is $\epsilon$.  
\par
\begin{figure}[t!]
    \centering
    \includegraphics[width=0.75\linewidth]{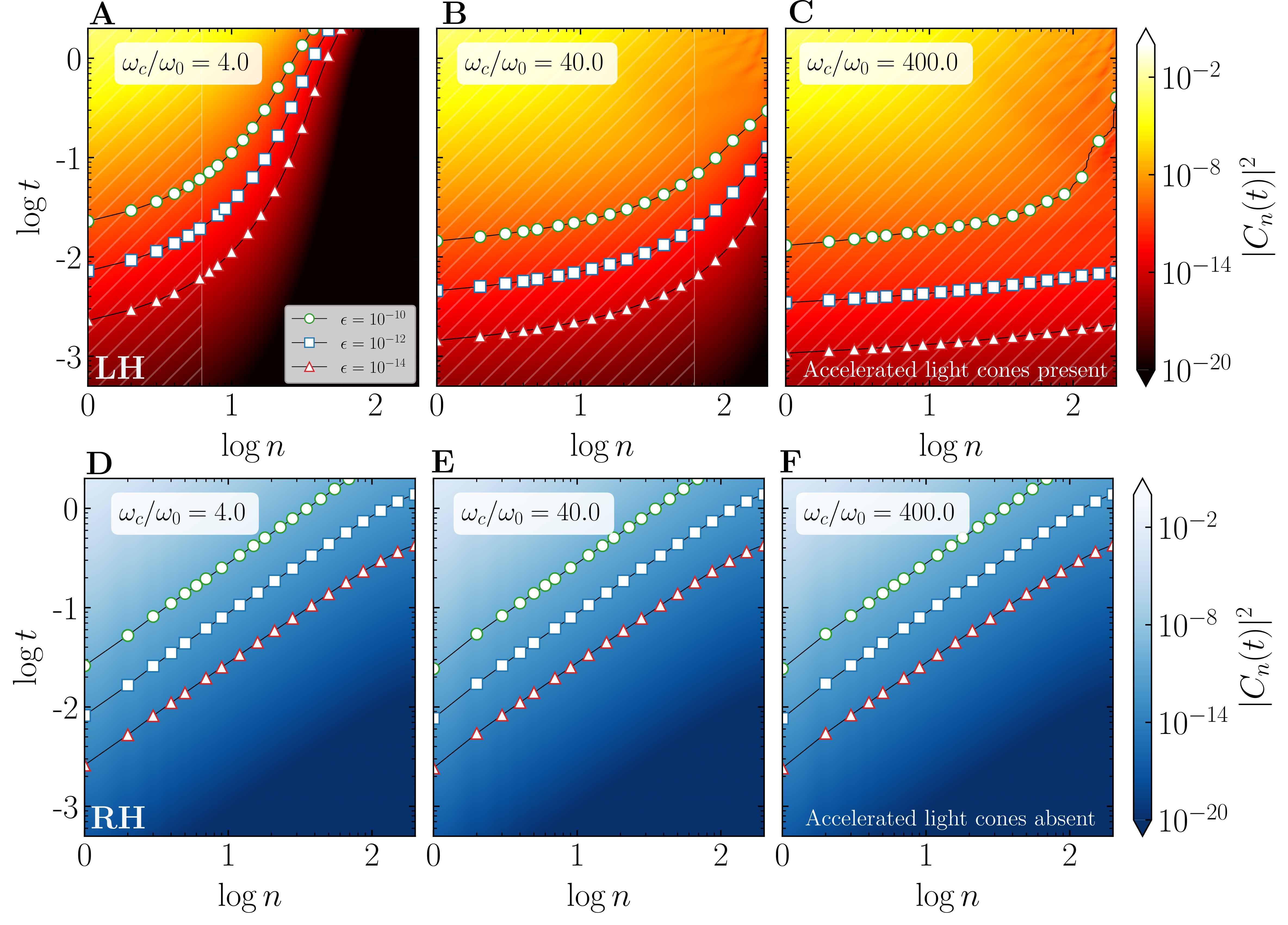}
    \caption{\textbf{Light cones in LHTL and RHTL.} Simulated photon intensity $|C_n(t)|^2$ (method described in Sec.~\ref{sec: Numerical Diagonalization}) along the transmission lines at different $\omega_c/\omega_0$ values, representing increasing $n_\star$ values ($N = 4001$, $\Delta/\omega_0 = 1$, $g/\omega_0 = 0.08$). The solid white markers, circle ($10^{-10}$), square ($10^{-12}$) and triangle ($10^{-14}$) denote different $\epsilon$-cones, and the solid black lines are guides to the eye.}
    \label{fig: Light-cones}
\end{figure}

For the LHTL, the running-exponent analysis gives
\begin{equation}
    \gamma^{\rm LH}(z)= z\,\frac{K_{1}(z)}{2K_{0}(z)},
\end{equation}
which increases with distance and crosses $\gamma=1$ at $z=z_2 \approx 1.55$, corresponding to $\alpha(z_2)=2$. Hence, $\gamma^{\rm LH}(z)<1$ for $z<z_2$, or equivalently $n<z_2 n_\star$, implies accelerated light-cones in this window, as shown in Fig.~\ref{fig: Light-cones}A--C. For $z>z_2$, the hopping becomes effectively short-range and the propagation crosses over to an ordinary linear cone.
\par
For the RHTL, by contrast, $\alpha^{\rm RH}(z)=2$ at all distances, so that
\begin{equation}
    \gamma^{\rm RH}(z)=1,
\end{equation}
yielding linear light-cones throughout, consistent with Fig.~\ref{fig: Light-cones} D--F.
Inclusion of momentum-dependence of coupling suppresses the light cone exponent of LHTL further, whereas it enhances the light cone exponent in RHTL leading to deceleration w.r.t. linear propagation.
\begin{equation}\label{eq: Exact gk LH light-cone}
    t^{\rm LH}_\epsilon(n) \sim \frac{\epsilon}{n^{1/4}K_{1/4}(n/n_\star)}, \quad t^{\rm RH}_\epsilon(n) \sim n^{3/2},
\end{equation}
%

%
\section{Numerical Diagonalization}\label{sec: Numerical Diagonalization}
%
\subsection{Bound state}
To complement the continuum analysis, we also compute the qubit–photon bound-state profile by exact diagonalization of the Hamiltonian for a finite waveguide with $N$ modes. The Hamiltonian we diagonalize is, 
\begin{equation}
    H \;=\; \Delta \,\sigma^+\sigma^- 
    + \sum_k \omega_k \, a_k^\dagger a_k
    + \sum_k \bigl( g_k \,\sigma^+ a_k + g_k^\ast \, a_k^\dagger \sigma^- \bigr),
    \label{eq:H_waveguide_singleexc}
\end{equation}
where $\Delta$ is the bare qubit frequency, $\omega_k$ is the dispersion relation of the transmission line, and $g_k$ is the momentum-dependent coupling we derived in earlier sections. In the single-excitation sector we introduce the operator vector $\Phi = \left( \sigma^-, a_{k_1}, a_{k_2}, \dots, a_{k_N} \right)^T$, so that the Hamiltonian can be compactly written as $H = \Phi^\dagger h(k) \Phi$, where 
\begin{equation}
    h(k) = \begin{pmatrix}
        \Delta & g_{k_1} & g_{k_2} & \dots & g_{k_N}\\
        g^*_{k_1} & \omega_{k_1} & 0 & \dots & 0\\
        g^*_{k_2} & 0 & \omega_{k_2} &  \dots & 0\\
        \vdots & \vdots &\vdots &\ddots & \vdots\\
        g^*_{k_N} & 0 & \dots & \dots & \omega_{k_N}
     \end{pmatrix}
\end{equation}
The $k$-space Hamiltonian $h(k)$ is then diagonalized numerically to get $N+1$ eigenvalues $E_\nu$ and their corresponding eigenvectors,
\begin{equation}
    |\psi_{(\nu)}\rangle = C_e^{(\nu)} |e,0\rangle + \sum_k C_k^{(\nu)}|g,1_k\rangle
\end{equation}
The bound states are identified by their eigenvalues $E_{\nu}$ lying outside the range $[\omega_\text{IR},\;\omega_\text{UV}]$. For each of those eigenvalues, the photonic coefficients $C_k^{(\nu)}$ are stored and the real-space bound-state profile amplitudes are given as $\phi_n^{(\nu)} \equiv \langle g,1_n|\psi_{(\nu)}\rangle$, where $\langle g,1_n| = \langle g,0| a_n$ and $a^\dagger_n = N^{-1}\sum_k e^{-ikn}a^\dagger_k$, is therefore given as discrete Fourier transform,
\begin{equation}\label{SMeq: DFT for photon envelope}
    \phi_n^{(\nu)} = \frac{1}{\sqrt{N}} \sum_k e^{ikn}\; C_k^{(\nu)}.
\end{equation}
%
\subsection{Scattering state}\label{subsec: Numerical Diagonalization scattering-state}
%
To compute the real-time dynamics in a finite waveguide, we evolve an initially excited emitter state, $ |\psi(0)\rangle = (1,0,0,\dots)^T $, under the single-excitation Hamiltonian according to
\[
|\psi(t)\rangle = e^{-iHt}|\psi(0)\rangle.
\]
For numerical evaluation, we diagonalize $H$ to obtain its eigenvalues $E_\nu$ and eigenvectors collected in a unitary matrix $V$, such that
\[
V^\dagger H V = \mathrm{diag}(E_0,E_1,\dots,E_N) \equiv E.
\]
In this basis, each eigenmode evolves independently with phase $e^{-iE_\nu t}$. Accordingly, the eigenstates may be written as
\begin{equation}
    |E_j\rangle = \sum_{m=0}^{N} V_{mj}\,|m\rangle,
\end{equation}
where $|0\rangle \equiv |e,0\rangle = |\psi(0)\rangle$ denotes the initially excited emitter state, and $|k\rangle \equiv |g,1_k\rangle$ labels the single-photon waveguide states. The full time evolution is therefore
\begin{equation}
    |\psi(t)\rangle
    =
    e^{-iHt}|\psi(0)\rangle
    =
    V e^{-iEt} V^\dagger |\psi(0)\rangle.
\end{equation}
Projecting back onto the original basis gives
\begin{equation}
    \psi_m(t)
    \equiv
    \langle m|\psi(t)\rangle
    =
    \sum_{j=0}^{N} V_{mj}V_{0j}^*\,e^{-iE_j t}.
\end{equation}
The emitter amplitude and momentum-space photonic amplitudes are then given by the first and remaining components of $\psi_m(t)$, respectively,
\begin{equation}
    C_e(t)\equiv \psi_0(t)=\sum_{j=0}^{N}|V_{0j}|^2 e^{-iE_j t},
    \qquad
    C_k(t)\equiv \psi_k(t)=\sum_{j=0}^{N}V_{kj}V_{0j}^* e^{-iE_j t}.
\end{equation}
Finally, the real-space photonic field is obtained by Fourier transforming the momentum-space amplitudes,
\begin{equation}
    \phi_n(t)=\frac{1}{\sqrt{N}}\sum_k e^{ikn}\,C_k(t).
\end{equation}